\documentclass[11pt,a4paper,english,twoside]{article}

\usepackage{a4wide}
\usepackage{amssymb, amsmath, amsthm}
\usepackage{graphicx}
\usepackage{subcaption}
\usepackage[all]{xy}
\usepackage{enumerate}
\usepackage[pdftex,hyperref,svgnames]{xcolor}
\usepackage[pdftex,colorlinks=true,
pdfstartview=FitV,
pdfnewwindow=true,
linktoc = page,
linkcolor= Red,
citecolor= blue,
urlcolor= blue,
hyperindex=true,
hyperfigures=false]{hyperref}
\hypersetup{linktocpage}
\usepackage{dsfont}
\usepackage{empheq}
\usepackage{cite}
\usepackage{float}
\usepackage{cancel}
\usepackage{relsize}
\usepackage{soul}
\usepackage{enumitem}
\usepackage{hhline}

\usepackage{graphicx}
\newcommand{\plus}{\raisebox{.25\height}{\scalebox{.6}{+}}}

\newcommand{\beq}{\begin{equation}}
\newcommand{\eeq}{\end{equation}}
\def\bea#1\eea{\begin{align}#1\end{align}}
\def\beal#1\eeal{\begin{subequations}\begin{align}#1\end{align}\end{subequations}}
\newcommand{\nn}{\nonumber}

\newcommand{\R}{\mathcal{R}}

\def\del {\partial}
\def\d {{\rm d}}
\def\mmm {\mathcal{M}}

\newcommand{\f}[2]{f^{#1}{}_{#2}}
\newcommand{\g}{\mathfrak{g}}

\begin{document}
\numberwithin{equation}{section}

\begin{titlepage}

\begin{center}

\phantom{DRAFT}

\vspace{1.2cm}

{\LARGE \bf{Exploring the landscape of (anti-) de Sitter\vspace{0.3cm}\\ and Minkowski solutions: group manifolds,\vspace{0.4cm}\\ stability and scale separation}}\\

\vspace{2.2 cm} {\Large David Andriot$^{1}$, Ludwig Horer$^{2}$, Paul Marconnet$^{3}$}\\
\vspace{0.9 cm} {\small\slshape $^1$ Laboratoire d’Annecy-le-Vieux de Physique Th\'eorique (LAPTh),\\
UMR 5108, CNRS, Universit\'e Savoie Mont Blanc (USMB),\\
9 Chemin de Bellevue, 74940 Annecy, France}\\
\vspace{0.2 cm} {\small\slshape $^2$ Institute for Theoretical Physics, TU Wien\\
Wiedner Hauptstrasse 8-10/136, A-1040 Vienna, Austria}\\
\vspace{0.2 cm} {\small\slshape $^3$ Institut de Physique des 2 Infinis de Lyon\\
Universit\'{e} de Lyon, UCBL, UMR 5822, CNRS/IN2P3\\
4 rue Enrico Fermi, 69622 Villeurbanne Cedex, France}\\
\vspace{0.5cm} {\upshape\ttfamily andriot@lapth.cnrs.fr; ludwig.horer@tuwien.ac.at;\\
marconnet@ipnl.in2p3.fr}\\

\vspace{2.8cm}

{\bf Abstract}
\vspace{0.1cm}
\end{center}

\begin{quotation}
We classified in \cite{Andriot:2022way} certain 10d supergravity solutions with a 4d de Sitter, Minkowski or anti-de Sitter spacetime. We then found new solutions in previously unexplored classes. In this paper we study their properties, compare them to swampland conjectures, and make new observations.

Using new numerical tools, we first identify all Lie algebras underlying the 6d group manifolds, allowing us to discuss their compactness. We then investigate scale separation, and prove related no-go theorems. Last but not least, we automatize and analyze the stability of all solutions. This leads us to propose the Massless Minkowski Conjecture, claiming the systematic presence of a 4d massless scalar field.
\end{quotation}

\end{titlepage}

\newpage

\tableofcontents

\section{Introduction and main results}

String theory backgrounds with a maximally symmetric spacetime are central in many research questions. De Sitter, Minkowski or anti-de Sitter backgrounds appear in a wide range of topics going from phenomenology to holography and quantum field theory. Part of this variety is captured by the swampland program \cite{Vafa:2005ui, Brennan:2017rbf, Palti:2019pca}, which aims at characterising outcomes of quantum gravity theories, such as string theory. Consequently, backgrounds with maximally symmetric spacetimes are subject to several conjectures regarding their existence and properties, currently under scrutiny. With these motivations in mind, we provided in a companion paper \cite{Andriot:2022way} a classification of certain 10d type IIA/B supergravity solutions with a maximally symmetric spacetime, which are candidates for classical and perturbative string backgrounds. This classification led us in particular to find new kinds of solutions. In this paper, we study various properties of these solutions and compare them to swampland conjectures; this analysis will reveal some interesting novelties.

The solutions classified share an ansatz which is common in the literature. It typically allows for a consistent truncation towards 4d gauged supergravities \cite{Andriot:2019wrs}. The 10-dimensional (10d) spacetime is a direct product of a 4d maximally symmetric spacetime and a 6d group manifold. In a certain basis, flux components are constant, as well as $D_p$-brane and orientifold $O_p$-plane contributions, which are smeared. The work of \cite{Andriot:2022way} classifies such solutions having non-zero source contributions appearing in the Bianchi identities, sometimes referred to as having a non-vanishing tadpole. Many well-known solutions enter this classification: the de Sitter ones of \cite{Danielsson:2011au, Andriot:2020wpp}, the Minkowski ones of \cite{Giddings:2001yu, Grana:2006kf}, the anti-de Sitter ones of \cite{DeWolfe:2005uu, Camara:2005dc, Caviezel:2008ik}, to cite just a few; a complete list is provided in \cite{Andriot:2022way}.

This classification led us to two important results. First, we could search and find new solutions in previously unexplored solution classes: for instance, we found de Sitter solutions with $O_4$ and $O_6$ (class $m_{46}$), Minkowski solutions with $O_5$ along 3 different direction sets (class $s_{555}$), or anti-de Sitter solutions with $O_5$ along 2 different sets (class $s_{55}$). Second, the classification and searches gave us an overview of the possible solutions, and it led us to Conjecture 4 on de Sitter solutions: those need at least 3 intersecting sets of $O_p/D_p$, which means that they can only be found in corresponding 4d theories having at most ${\cal N}=1$ supersymmetry.

In this paper, we study these new solutions on several important aspects, allowing us to compare them to related swampland conjectures. We develop on the way methods and numerical tools, that we describe and make available. We present in the following three different aspects and the main results obtained for each of them.

For completeness, let us add that we do not include in our solution ansatz other ingredients a priori allowed in 10d supergravities: $N\!S_5$-branes, Kaluza--Klein monopoles (KKm) or anti-$D_p$-branes ($\overline{D}_p$). KKm look promising for de Sitter solutions, see e.g.~\cite{Silverstein:2007ac, Blaback:2018hdo, Blaback:2019zig}. The corresponding sourced Bianchi identity is a violation of Jacobi identities or Riemann Bianchi identity \cite{Villadoro:2007yq, Andriot:2014uda}, and may lead to a tadpole. On a compact manifold, one may wonder whether this would require the analogue of an orientifold, sometimes denoted KKO \cite{Villadoro:2007yq, Dibitetto:2019odu}. Having a localized 10d solution with a KKm, possibly a KKO, \textsl{together} with $O_p/D_p$ seems to us so far out of sight (localized $O_p/D_p$ solutions can already be very difficult to obtain); the same goes for $N\!S_5$-branes. So for simplicity we do not include any of those NSNS objects in our ansatz. The $\overline{D}_p$ have been argued to help regarding the stability of de Sitter solutions \cite{Kallosh:2018nrk} (without however finding a 10d supergravity solution realising this idea). But $\overline{D}_p$ have the disadvantage of being unstable in presence of a $D_p$, so one has to make sure that they are not along the same directions. We thus focus on the simpler setting with only $O_p/D_p$ as sources. Note also that the classification made in \cite{Andriot:2022way} does not restrict to supersymmetry-preserving $O_p/D_p$ configurations; rather the amount of preserved supersymmetry is determined a posteriori for each solution class in \cite[Sec. 2.4.2]{Andriot:2022way}.

\subsection*{Group manifolds}

The supergravity solutions found on 4d maximally symmetric spacetimes admit as extra dimensions a 6d group manifold $\mmm$. The latter is encoded in structure constants $f^a{}_{bc}$ of an underlying Lie algebra $\mathfrak{g}$. Contrary to previous approaches \cite{Grana:2006kf, Danielsson:2011au}, the search for solutions performed in \cite{Andriot:2020wpp,Andriot:2021rdy,Andriot:2022way} does not fix this algebra from the start, but leaves the structure constants free as variables; they are still bound to verify the Jacobi identities, and the orientifold projections. This approach provides more freedom in the search for new solutions, but it has the drawback that the algebra, and subsequent group manifold $\mmm$, have to be identified a posteriori. In particular, the compactness of $\mmm$ is a priori not guaranteed. The group manifold will be compact if the algebra is, or if one can find a discrete subgroup that makes the group compact after quotienting, a.k.a.~a lattice (see \cite{Bock, Andriot:2010ju}). Most of the time, this can only be settled once the algebra $\mathfrak{g}$ is identified. This identification is also needed to study whether a solution can be a classical string background, as done e.g.~in \cite{Andriot:2020vlg}; we do not perform this further analysis here, but hope to come back to it in future work. In Section \ref{sec:algcomp}, we thus tackle this task of identifying the algebras of the new solutions found, having only their structure constants. We develop a method, and the numerical tools {\tt AlgId} and {\tt AlgIso}. This eventually allows us to \textsl{identify algebras of all solutions}. We present these results and discuss the compactness of $\mmm$ in Section \ref{sec:resultsalg}. We prove this way that the de Sitter solution $s_{55}^+$ 19 of \cite{Andriot:2021rdy} is on a compact manifold, a point not previously established. Being only mildly unstable (see Section \ref{sec:stab}), this solution is even more interesting.

In more detail, let us mention that there exist 100 (isomorphism classes of) real 6d indecomposable unimodular solvable Lie algebras, classified in tables \cite{Bock, mathbook}, to which one should add the decomposable ones, and the 16 real 6d unimodular non-solvable Lie algebras. These numbers imply that identifying the algebra, simply with the structure constants, is at first sight not trivial. In addition, the structure constants in our solutions have been obtained in a certain basis of the algebra; typically an isomorphism or change of basis is required to match the algebra as listed in known tables. This makes the identification even more difficult. Our tools and method make use of basis-invariant properties to help with this identification. Let us stress once again that this way, algebras of all solutions are eventually identified, and a compact manifold is found in almost all solution classes, so the difficulties raised by this approach can be considered as overcome.

\subsection*{Stability}

The stability of solutions with maximally symmetric spacetime is the topic of several swampland conjectures, claiming in particular instability for non-supersymmetric backgrounds: see for instance \cite{Andriot:2018wzk, Garg:2018reu, Ooguri:2018wrx, Andriot:2018mav} for de Sitter, \cite{Acharya:2019mcu, Acharya:2020hsc} for Minkowski and \cite{Ooguri:2016pdq} for anti-de Sitter solutions. In Section \ref{sec:stab}, we study the perturbative stability of the new solutions found in \cite{Andriot:2022way}, which are likely to be non-supersymmetric (see Section \ref{sec:stabMink}). Since these solutions were found in previously unexplored solution classes, they could in principle exhibit new physics that would contradict expectations formulated in the swampland conjectures; we make this comparison in Section \ref{sec:resultstab} when presenting our results on the stability of all these solutions. We find in particular the anti-de Sitter solutions $s_{55}^-$ 2,3,4 to be perturbatively stable (in the fields considered), motivating further study in view of the conjecture of \cite{Ooguri:2016pdq}. We propose also a new conjecture for Minkowski solutions, detailed below.

We first consider certain 4d scalar fluctuations around our solutions, then determine their mass spectrum and read from it the perturbative stability of the solutions. More precisely, these fluctuations are governed by a 4d effective action of the form
\begin{equation}
{\cal S} = \int \d^4 x \sqrt{|g_4|} \left(\frac{M_p^2}{2} \R_4 - \frac{1}{2} g_{ij} \del_{\mu}\phi^i \del^{\mu}\phi^j - V \right) \ ,\label{S4dgen}
\end{equation}
where $g_{ij}$ is the field space metric, $V$ a scalar potential for the scalar fields $\phi^i$, and $M_p$ the 4d reduced Planck mass, given by $M_p^2 = \frac{1}{\kappa_{10}^2} \int \d^6 y \sqrt{|g_6|}\ g_s^{-2} $. This action is obtained after dimensional reduction from the 10d type II supergravities, as a consistent truncation. Our 10d solutions correspond to critical points of the 4d potential, $\del_{\phi^i} V =0$, with the cosmological constant given at this point by $\Lambda = \frac{V}{M_p^{2}} = \frac{1}{4} {\cal R}_4 $. The perturbative stability of our solutions can be read from the 4d mass spectrum, given by the eigenvalues of the mass matrix $M$, with $M^i{}_j = g^{ik} \nabla_{\!\phi^k} \del_{\phi^j} V$, at the critical point.

The set of 4d scalar fields $\phi^i$ considered will be restricted to $(\rho,\tau,\sigma_I)$, with $\rho$ the (6d) volume, $\tau$ the 4d dilaton, and $\sigma_I$ related to internal volumes wrapped by each $O_p/D_p$ source set $I$. This is motivated by the well-verified proposal of \cite{Danielsson:2012et}. It states that the tachyon systematically observed in 10d supergravity de Sitter solutions (see e.g.~\cite{Andriot:2021rdy}) lies only among these fields. We then consider these fields, and for each source configuration of each solution class, we need to provide the kinetic terms in the action \eqref{S4dgen}, i.e.~the field space metric $g_{ij}$, and the scalar potential $V(\rho, \tau, \sigma_I)$. We do so building on \cite{Hertzberg:2007wc, Silverstein:2007ac, Danielsson:2012et, Andriot:2018ept, Andriot:2019wrs, Andriot:2020lea}. One difficulty to overcome is a possible redundancy among the fields $\sigma_I$, when the internal volumes wrapped by the sources are not independent. Redundant fields should then be identified and removed through a field redefinition. \textsl{All these tasks have been automatized} in the numerical tool {\tt MaxSymSolSpec}: after identifying independent fields, computing $g_{ij}$ and $V$, it determines the mass spectrum for a given solution. We obtain this data for all solutions of \cite{Andriot:2022way}, and determine this way their stability: see Section \ref{sec:resultstab}. The complete stability data will be provided for each solution in Appendix C of the revised version of \cite{Andriot:2022way}.

As for de Sitter solutions in \cite{Andriot:2022way}, the overview of our Minkowski solutions allows us here to formulate in \eqref{conjMink} and \eqref{conjMinkswamp} a conjecture related to their stability:
\bea
& \text{{\bf Massless Minkowski Conjecture:}}\\
& \nn\\[-10pt]
& \text{\textsl{10d supergravity solutions compactified to 4d Minkowski always admit a 4d massless scalar,}}\nn\\
& \text{\textsl{among the fields $(\rho,\tau,\sigma_I)$.}}\nn
\eea
We refer to Section \ref{sec:stabMink} and Appendix \ref{ap:spec} (mass spectra) for support and discussion of this conjecture. In comparison to previous related statements, let us emphasize the specification of the fields $(\rho,\tau,\sigma_I)$, and the fact that the claim does not depend on supersymmetry (neither that of the solution, nor of the 4d theory). These two points directly hint at a possible relation between this massless mode in Minkowski and the de Sitter tachyon. Let us also emphasize that the conjectured massless scalar field is not necessarily a flat direction. It would be interesting to investigate whether the (non-perturbative) instability, conjectured for non-supersymmetric Ricci flat compactifications to Minkowski in \cite{Acharya:2019mcu, Acharya:2020hsc}, has any relation to the massless scalar mentioned here. The above conjecture is also reminiscent of the tadpole conjecture \cite{Bena:2020xrh}, however, with several differences. Last but not least, we discuss in Section \ref{sec:stabMink} the possibility of a swampland corollary, together with a strong version of the conjecture. In the latter, we propose in addition the absence of a 4d tachyon in a Minkowski solution. This implies that the inequalities of the refined de Sitter conjectures of \cite{Garg:2018reu,Ooguri:2018wrx,Andriot:2018mav} are saturated, meaning
\beq
0 = {\rm min}\ \nabla \del V = \frac{ V}{{M_p}^2} = \frac{|\nabla V|}{M_p} \ .
\eeq

\subsection*{Scale separation}

In Section \ref{sec:scalesep}, we study the question of scale separation in the solutions found in \cite{Andriot:2022way}. Having scale separation in 4d is the requirement that energy scales associated to towers of modes, for instance the first non-zero mass in a scalar Kaluza--Klein tower, is much higher than a typical 4d effective theory energy scale, for example that of a non-zero cosmological constant. Having such a separation of scales is needed for a low energy truncation, i.e.~having a 4d effective theory while infinite towers of modes are above a cut-off energy. For anti-de Sitter solutions, where the matter is mostly discussed, this is expressed by $m^2/|\Lambda| \gg 1$, where $m$ typically refers to the first non-zero mass of a tower. The same requirement can be made for de Sitter solutions, and was for instance discussed in \cite{Andriot:2020vlg}. For Minkowski solutions where $\Lambda=0$, one would rather require a mass gap, between light modes whose mass is typically set by the scalar potential, and the first massive mode of a tower, e.g.~the first massive eigenmode of the Laplacian operator. In Section \ref{sec:scalesep}, we focus on the anti-de Sitter and Minkowski solutions of \cite{Andriot:2022way}: since they were found in new solution classes, whether they exhibit scale separation is unknown and should be investigated.

While being an old topic (see e.g.~\cite{Caviezel:2008ik, Tsimpis:2012tu, Petrini:2013ika, Gautason:2015tig}), scale separation in anti-de Sitter solutions has received renewed attention following swampland conjectures on the topic \cite{Gautason:2018gln, Lust:2019zwm, Blumenhagen:2019vgj, Font:2019uva, Buratti:2020kda, Cribiori:2022trc}, leading to many recent works, for instance on concrete anti-de Sitter solutions \cite{Junghans:2020acz, Marchesano:2020qvg, Farakos:2020phe, Apers:2022zjx, Emelin:2022cac}. Of particular interest are the so-called DGKT solutions \cite{DeWolfe:2005uu, Camara:2005dc, Acharya:2006ne}: 10d supergravity solutions with 4d anti-de Sitter spacetime (and smeared sources) that fall, at least for some of them, in our solution class $s_{6666}$. These solutions were classified in \cite{Marchesano:2019hfb} and new concrete examples were found in \cite{Cribiori:2021djm}. Not only do these solutions exhibit scale separation, they do so with parametric control, meaning that the separation can be tuned thanks to a parameter. It is important to distinguish the two concepts, as stressed in \cite{Andriot:2020vlg} for de Sitter solutions (see also \cite{Cicoli:2021fsd}): one can simply look for a satisfactory hierarchy of scales in a given solution, without asking for this hierarchy to be tunable. Another concept is that of classicality of a 10d supergravity solution, namely whether it belongs to the classical regime of string theory. DGKT solutions have the property that they can be classical, with the same parametric control. On the contrary, anti-de Sitter solutions exhibiting scale separation and belonging to the class $m_{5577}$ \cite{Caviezel:2008ik, Petrini:2013ika} cannot be made classical \cite{Cribiori:2021djm}. We will not study classicality here, but it is interesting to note such a difference between two seemingly close solution classes.

Scale-separated solutions of 10d supergravity with a 4d anti-de Sitter spacetime have been found in $s_{6666}$ on a 6d torus, or in $m_{5577}$ on a 6d nilmanifold; the two settings are at first sight T-dual. There are reasons to believe that these two geometries, together with manifolds with a Ricci flat metric, are preferred among group manifolds to achieve scale separation. First, it has been argued \cite[Sec. III]{Andriot:2018wzk} that other group manifolds have structure constants (i.e.~spin connection components) leading to energy scales higher than the first massive Kaluza--Klein scale. These structure constants, giving the curvature of some internal subspaces, cannot be truncated for the solution to exist; in particular they must be present in the 4d theory if one wants to recover there the 10d solution as a critical point. Therefore they prevent from achieving scale separation. A second argument is that if the Kaluza--Klein scale is comparable to the 6d curvature ${\cal R}_6$, then scale separation cannot be achieved \cite{Gautason:2015tig}. As shown however in \cite{Andriot:2018tmb} and \cite[Foot. 8]{Andriot:2019wrs}, nilmanifolds precisely allow for a gap between ${\cal R}_6$ and the Kaluza--Klein scale; this is obviously true for a Ricci flat manifold. In Section \ref{sec:nogo}, we then study the possibility of finding anti-de Sitter solutions in the new solution classes $s_{55}$ and $m_{46}$ on nilmanifolds or Ricci flat ones, and \textsl{we conclude negatively with no-go theorems}. This hints at an \textsl{absence of scale separation}; a comparison is made to the seemingly close solution classes $s_{6666}$ and $m_{5577}$. The presence of internal directions with $D_p$ but no $O_p$ appears as a key difference.

\section{Algebra identification and compactness}\label{sec:algcomp}

In this section, we focus on the identification of Lie algebras $\mathfrak{g}$ underlying the 6d group manifolds $\mmm$ in the solutions found, and the compactness of $\mmm$. We first recall in Section \ref{sec:algrecap} a few useful elements of algebras. We illustrate those in various examples in Section \ref{sec:algex}. We then present in Section \ref{sec:mettools} the method used for this identification, as well as the (partly numerical) tools developed. We finally present our results in Section \ref{sec:resultsalg}, namely the identification of all 6d algebras appearing in the solutions found in \cite{Andriot:2020wpp,Andriot:2021rdy,Andriot:2022way}, and what can be said on the compactness of the corresponding group manifold $\mmm$.

\subsection{Elements of algebra}\label{sec:algrecap}

We consider a real 6d Lie algebra $\mathfrak{g}$. In a given basis, it is expressed by the commutation relations of the 6 vectors $\left\{ E_a \right\}$, $a=1,...,6$, in terms of the structure constants $f^a{}_{bc}$
\begin{equation}
\mathfrak{g} \; : \; \left[E_b,E_c \right] = \f{a}{bc}\, E_a \,,
\end{equation}
and the structure constants are bound to verify the Jacobi identities. Real 6d Lie algebras are classified. Levi's decomposition indicates that any Lie algebra $\g$ is the semi-direct sum of a semi-simple algebra $\mathfrak{s}$ and a solvable ideal $\mathfrak{r}$ called the radical of $\g$
\begin{equation}
\g = \mathfrak{s} \ \plus \!\!\!\!\! \supset \mathfrak{r} \,.
\end{equation}
We will distinguish two cases:
\begin{itemize}
\item $\g=\mathfrak{r}$ is solvable (to be defined below).
\item $\g$ is not solvable. This last case can be further divided in three situations: $\g=\mathfrak{s}$ is semi-simple, $\g = \mathfrak{s} \oplus \mathfrak{r}$ is a \textit{direct} sum, or $\g = \mathfrak{s} \ \plus \!\!\!\!\! \supset \mathfrak{r}$ is a (non-trivial) \textit{semi-direct} sum. This division will however not be crucial to us since we will only consider few algebras that are not solvable.
\end{itemize}

We now introduce a few elements to define solvability, and a particular case, nilpotency, of a Lie algebra; a mathematical review on solvable algebras and the corresponding solvmanifolds can be found in \cite{Andriot:2010ju}. We first recall that an ideal $\mathfrak{i}$ of an algebra $\mathfrak{g}$ is a subalgebra that verifies $\left[ \mathfrak{g}, \mathfrak{i} \right] \subseteq \mathfrak{i}$. Any Lie algebra $\g$ possesses three series of ideals: the derived series, the lower central series, and the upper central series. We will only need the first two: the lower central series $\left\{\g_{(k)}\right\}_{k \in \mathbb{N}}$ is defined recursively as follows
\begin{equation}
\g_{(k)} = \left[\g_{(k-1)}, \g \right] \,, \quad \g_{(0)} = \g \,,
\end{equation}
while the derived series $\left\{\g^{(k)}\right\}_{k \in \mathbb{N}}$ is defined as
\begin{equation}
\g^{(k)} = \left[\g^{(k-1)}, \g^{(k-1)} \right] \,, \quad \g^{(0)} = \g \,.
\end{equation}
The sets of successive dimensions of ideals in the lower central and derived series are respectively denoted CS and DS. These two sets of integers are readily computable from the list of structure constants, and are often different from one Lie algebra to another one. They are basis independent: they will then be useful for the identification of algebras. And indeed, a first use can be seen through the following definitions. An algebra is solvable iff $\exists k$ s.t. $\g^{(k)}=0$; in other words, its DS ends with 0. An algebra is nilpotent iff $\exists k$ s.t. $\g_{(k)}=0$; in other words, its CS reaches 0. The latter is a particular case of the former.

Another useful definition is that of the nilradical $\mathfrak{n}$ of a solvable Lie algebra $\g$: $\mathfrak{n}$ is the maximal nilpotent ideal of $\g$. It is unique, and solvable algebras are classified according to their nilradical. Determining it is thus an important step towards identifying a solvable algebra.

A necessary condition for compactness of the group manifold is the unimodularity (also known as unipotence) of the Lie algebra: this is defined as
\begin{equation}
\label{unip}
\forall b,\quad \sum_a \f{a}{ab} = 0 \,.
\end{equation}
The ansatz used to find solutions required a stronger condition, namely $\f{a}{ab} = 0$ without sum on $a$. This is motivated by an appropriate choice of basis \cite{Andriot:2010ju, Andriot:2022way}. The list of (isomorphism classes of) indecomposable unimodular real solvable Lie algebras up to dimension 6 is given in \cite{Bock}; there are 100 6-dimensional ones. The list of unimodular real non-solvable 6d Lie algebras is given in \cite{Danielsson:2011au} and below in Table \ref{tab:alg}: there are only 16 of them. We will also make use of the classification given in \cite{mathbook}, which does not restrict to unimodular algebras, but gives the CS and DS values for all algebras.

A last element which will be useful is the Killing form. It is a symmetric bilinear form on the algebra $\mathfrak{g}$
\begin{equation}
B(x,y) = \text{Tr}\left(\text{ad}(x) \cdot \text{ad}(y)\right) \,, \quad x,y \in \mathfrak{g} \qquad \Longleftrightarrow \qquad B_{ab} = \f{c}{da} \, \f{d}{cb} \,. \label{killing}
\end{equation}
This (0,2)-tensor, equivalently represented by a (symmetric) matrix, has interesting properties. To start with, one has
\begin{itemize}
\item $\mathfrak{g}$ is semi-simple iff $B$ has a non-zero determinant,
\item $\mathfrak{g}$ is solvable iff $B(\g, \left[\g,\g\right])=0$,
\item If $\mathfrak{g}$ is nilpotent then $B$ is identically zero.
\end{itemize}
In addition, the signature of $B$ is invariant under a real change of basis. This will be of interest to us for the identification of algebras: we will be interested in the number of positive and negative eigenvalues of $B$. Finally, if $B$ only has negative eigenvalues (implying that it is semi-simple) then $\g$ is compact.

A solvable algebra $\g$ gives rise to a compact group manifold $\mmm$ whenever a lattice can be found. Let us consider a discrete subgroup $\Gamma$ of the group $G$ associated to $\g$. This $\Gamma$ is a lattice if the quotient $G/\Gamma$ is compact. This quotient is then precisely the group manifold $\mmm$, and it is called a solvmanifold. A particular case is a nilmanifold, the quotient of a nilpotent group by a lattice. Given an indecomposable solvable algebra, whether or not a lattice can be found is not always settled. As a consequence, we cannot always conclude on the compactness of $\mmm$ once the algebra is identified. We refer to \cite{Bock,Andriot:2010ju} for more details on this matter of compactness.

Beyond the elements presented above, many more exist, with associated methods to help identifying Lie algebras given in terms of their structure constants. We can mention, among others, the upper central series and their dimensions (US), the number of generalized Casimir invariants, decomposability properties, etc. We refer the interested reader to \cite{mathbook}. The above will be enough for our purposes.

\subsection{Examples of algebras}\label{sec:algex}

Let us illustrate the previous definitions with a few examples. We start with low dimensional real unimodular Lie algebras. In 1 or 2 dimensions, there are only $\mathfrak{u}(1)$ and $2\, \mathfrak{u}(1)$. In 3 dimensions, there are six of them. We give them below in terms of their non-zero structure constants in some basis; the directions numbering, 123 or 456, is chosen for convenience
\bea
3\, \mathfrak{u}(1)\, & : \quad (f^a{}_{bc} = 0) \qquad \mbox{(nilpotent)} \\
{\rm Heis}_3\, & : \quad f^4{}_{56} = 1 \qquad \mbox{(nilpotent)} \nn\\
\mathfrak{g}_{3.5}^{0} = \mathfrak{iso}(2)\, & : \quad f^4{}_{56} = 1,\ f^5{}_{46} = -1 \qquad \mbox{(solvable)} \nn\\
\mathfrak{g}_{3.4}^{-1} = \mathfrak{iso}(1,1)\, & : \quad f^4{}_{56} = 1,\ f^5{}_{46} = 1\ \Leftrightarrow \ f^4{}_{46} = 1,\ f^5{}_{56} = -1 \qquad \mbox{(solvable)} \nn\\
\mathfrak{so}(3) = \mathfrak{su}(2) \, & : \quad f^1{}_{23} = 1,\ f^2{}_{31} = 1,\ f^3{}_{12} = 1 \qquad \mbox{(simple)} \nn\\
\mathfrak{so}(2,1) = \mathfrak{sl}(2,\mathbb{R}) \, & : \quad f^1{}_{23} = 1,\ f^2{}_{31} = 1,\ f^3{}_{12} = -1 \qquad \mbox{(simple)} \nn
\eea
Of those, only $\mathfrak{so}(2,1)$ does not lead to a compact group manifold. Others are either compact, or admit lattices giving compact group manifolds.

In 4, 5 or 6 dimensions, the only (semi)-simple real unimodular Lie algebra that is indecomposable, is the following 6-dimensional one
\bea
\mathfrak{so}(3,1) \, : \quad & f^1{}_{23} = 1,\ f^2{}_{31} = 1,\ f^3{}_{12} = 1 \ , f^1{}_{56} = -1,\ f^5{}_{61} = 1,\ f^6{}_{15} = 1 \\
& f^2{}_{46} = 1,\ f^4{}_{62} = -1,\ f^6{}_{24} = -1 \ , f^3{}_{45} = -1,\ f^4{}_{53} = 1,\ f^5{}_{34} = 1 \,.\nn
\eea
This algebra is not compact. From these ingredients, one can build all 6-dimensional real unimodular Lie algebras, that are not solvable. Following \cite{Danielsson:2011au, mathbook} (and notations of \cite{Bock}), we list them here in Table \ref{tab:alg}, and determine some of their properties. We provide below few more comments on them, before turning to solvable algebras in 4, 5 or 6 dimensions.

\begin{table}[H]
  \begin{center}
    \begin{tabular}{|c||c|c|c|c|c|}
    \hline
 & Algebra & $\mmm$ compactness & CS & DS & Eigenvalues \\
\hhline{=::=====}
 & $\mathfrak{so}(3,1)$ & $\times$ & 6 & 6 & 3+,3- \\
Semi-simple & $\mathfrak{so}(3) \oplus \mathfrak{so}(3)$ & $\checkmark$ & 6 & 6 & 6- \\
 & $\mathfrak{so}(3) \oplus \mathfrak{so}(2,1)$ & $\times$ & 6 & 6 & 2+,4- \\
 & $\mathfrak{so}(2,1) \oplus \mathfrak{so}(2,1)$ & $\times$ & 6 & 6 & 4+,2- \\
    \hline
 & $\mathfrak{so}(3) \oplus 3\, \mathfrak{u}(1)$ & $\checkmark$ & 6,3 & 6,3 & 3- \\
 & $\mathfrak{so}(3) \oplus {\rm Heis}_3$ & $\checkmark$ & 6,4,3 & 6,4,3 & 3- \\
 & $\mathfrak{so}(3) \oplus \mathfrak{g}_{3.5}^{0}$ & $\checkmark$ & 6,5 & 6,5,3 & 4- \\
Direct sum & $\mathfrak{so}(3) \oplus \mathfrak{g}_{3.4}^{-1}$ & $\checkmark$ & 6,5 & 6,5,3 & 1+,3- \\
simple $\oplus$ solvable & $\mathfrak{so}(2,1) \oplus 3\, \mathfrak{u}(1)$ & $\times$ & 6,3 & 6,3 & 2+,1- \\
 & $\mathfrak{so}(2,1) \oplus {\rm Heis}_3$ & $\times$ & 6,4,3 & 6,4,3 & 2+,1- \\
 & $\mathfrak{so}(2,1) \oplus \mathfrak{g}_{3.5}^{0}$ & $\times$ & 6,5 & 6,5,3 & 2+,2- \\
 & $\mathfrak{so}(2,1) \oplus \mathfrak{g}_{3.4}^{-1}$ & $\times$ & 6,5 & 6,5,3 & 3+,1- \\
   \hline
 & $\mathfrak{so}(3)\ \plus \!\!\!\!\! \supset 3\, \mathfrak{u}(1)$ & $\checkmark$ & 6 & 6 & 3- \\
Semi-direct sum & $\mathfrak{so}(2,1)\ \plus \!\!\!\!\! \supset 3\, \mathfrak{u}(1)$ & $\times$ & 6 & 6 & 2+,1- \\
simple $\plus \!\!\!\!\! \supset$ solvable & $\mathfrak{so}(2,1)\ \plus \!\!\!\!\! \supset 2\, \mathfrak{u}(1) \oplus  \mathfrak{u}(1)$ & $\times$ & 6,5 & 6,5 & 2+,1- \\
 & $\mathfrak{so}(2,1)\ \plus \!\!\!\!\! \supset {\rm Heis}_3$ & $\times$ & 6 & 6 & 2+,1- \\
   \hline
    \end{tabular}
     \caption{All 6-dimensional real unimodular Lie algebras, that are not solvable. The compactness is that of an associated group manifold $\mmm$, possibly thanks to a lattice. CS denotes the successive dimensions of the ideals in the lower central series, and DS those of the derived series. Eigenvalues with $p+,m-$ denotes that the Killing form admits $p$ positive eigenvalues and $m$ negative ones, the remaining $6-p-m$ eigenvalues being 0.}\label{tab:alg}
  \end{center}
\end{table}

For completeness, we give as follows the structure constants for the semi-direct sum algebras appearing in Table \ref{tab:alg}
\bea
\mathfrak{so}(3)\ \plus \!\!\!\!\! \supset 3\, \mathfrak{u}(1) \, & : \quad f^1{}_{23} = 1,\ f^2{}_{31} = 1,\ f^3{}_{12} = 1,\ f^4{}_{35} = -1,\ f^5{}_{34} = 1,\nn\\
&\ \ \quad f^4{}_{26} = 1,\ f^6{}_{24} = -1,\ f^5{}_{16} = -1,\ f^6{}_{15} = 1  \\
\mathfrak{so}(2,1)\ \plus \!\!\!\!\! \supset 3\, \mathfrak{u}(1) \, & : \quad f^1{}_{23} = -1,\ f^2{}_{31} = -1,\ f^3{}_{12} = 1,\ f^4{}_{15} = -1,\ f^5{}_{14} = -1,\nn\\
&\ \ \quad f^4{}_{26} = 1,\ f^6{}_{24} = 1,\ f^5{}_{36} = -1,\ f^6{}_{35} = 1  \\
\mathfrak{so}(2,1)\ \plus \!\!\!\!\! \supset 2\, \mathfrak{u}(1) \oplus  \mathfrak{u}(1)\, & : \quad f^1{}_{12} = 2,\ f^2{}_{13} = -1,\ f^3{}_{23} = 2 ,\ f^4{}_{24} = 1,\ f^5{}_{25} = -1,\nn\\
&\ \ \quad f^5{}_{14} = 1,\ f^4{}_{35} = 1 \\
\mathfrak{so}(2,1)\ \plus \!\!\!\!\! \supset {\rm Heis}_3 \, & : \quad f^1{}_{12} = 2,\ f^2{}_{13} = -1,\ f^3{}_{23} = 2 ,\ f^5{}_{25} = 1,\ f^6{}_{26} = -1,\nn\\
&\ \ \quad f^6{}_{15} = 1,\ f^5{}_{36} = 1,\ f^4{}_{56} = 1 \,. \nn
\eea
For the last two, there could be no basis where $f^a{}_{ac}=0$ without sum on $a$.

Let us comment on the compactness of $\mmm$ indicated in Table \ref{tab:alg}. Most of the time, its non-compactness is due to $\mathfrak{so}(2,1)$. The semi-direct sum $\mathfrak{b} \ \plus \!\!\!\!\! \supset \mathfrak{f}$ can be interpreted geometrically as leading for $\mmm$ to a fibration, where the fiber comes from $\mathfrak{f}$ and is over a base generated by $\mathfrak{b}$. Indeed, when one moves in $\mathfrak{b}$, there is a change on the elements of $\mathfrak{f}$. In addition, if the base in a fiber bundle is non-compact, the manifold is non-compact as well. We conclude that $\mathfrak{so}(2,1)\ \plus \!\!\!\!\! \supset \mathfrak{f}$ are non-compact.\\

There are many more real unimodular solvable Lie algebras, in 4, 5 or 6 dimensions. Let us first focus on 6d decomposable ones. Those are a direct sum of lower dimensional real unimodular solvable algebras. If one of those is a 4d or 5d indecomposable one, then the rest can only be $2\, \mathfrak{u}(1)$ or $\mathfrak{u}(1)$. Otherwise, either the 6d algebra is $6\, \mathfrak{u}(1)$ or it contains a 3d indecomposable real unimodular solvable algebra: we list the 6d algebras built in this way in Table \ref{tab:algsolv3d}, together with some properties.

\begin{table}[H]
  \begin{center}
    \begin{tabular}{|c|c|c|c|c|}
    \hline
Algebra & $\mmm$ compactness & CS & DS & Eigenvalues \\
\hhline{=====}
 $3\, \mathfrak{u}(1) \oplus 3\, \mathfrak{u}(1)$ & $\checkmark$ & 6,0 & 6,0 & 0 \\
 ${\rm Heis}_3 \oplus 3\, \mathfrak{u}(1)$ & $\checkmark$ & 6,1,0 & 6,1,0 & 0 \\
 $\mathfrak{g}_{3.5}^{0} \oplus 3\, \mathfrak{u}(1)$ & $\checkmark$ & 6,2 & 6,2,0 & 1- \\
 $\mathfrak{g}_{3.4}^{-1} \oplus 3\, \mathfrak{u}(1)$ & $\checkmark$ & 6,2 & 6,2,0 & 1+ \\
 ${\rm Heis}_3 \oplus {\rm Heis}_3$ & $\checkmark$ & 6,2,0 & 6,2,0 & 0 \\
 $\mathfrak{g}_{3.5}^{0} \oplus {\rm Heis}_3$ & $\checkmark$ & 6,3,2 & 6,3,0 & 1- \\
 $\mathfrak{g}_{3.4}^{-1} \oplus {\rm Heis}_3$ & $\checkmark$ & 6,3,2 & 6,3,0 & 1+ \\
 $\mathfrak{g}_{3.5}^{0} \oplus \mathfrak{g}_{3.5}^{0}$ & $\checkmark$ & 6,4 & 6,4,0 & 2- \\
 $\mathfrak{g}_{3.4}^{-1} \oplus \mathfrak{g}_{3.5}^{0}$ & $\checkmark$ & 6,4 & 6,4,0 & 1+,1- \\
 $\mathfrak{g}_{3.4}^{-1} \oplus \mathfrak{g}_{3.4}^{-1}$ & $\checkmark$ & 6,4 & 6,4,0 & 2+ \\
   \hline
    \end{tabular}
     \caption{6-dimensional real unimodular solvable Lie algebras, that are decomposable and do not contain a 4d or 5d indecomposable subalgebra; rather they are a direct sum of 3-dimensional subalgebras. We also give some properties: the compactness of an associated group manifold $\mmm$ (thanks to a lattice), the CS and DS, and the number of positive and negative eigenvalues of the Killing form.}\label{tab:algsolv3d}
  \end{center}
\end{table}

Finally, real 6d unimodular solvable Lie algebras that are indecomposable are classified according to their nilradical as e.g.~in \cite{Bock}, into so-called ``isomorphism classes'': for any such algebra, one can find an isomorphism or change of basis that maps it to one (and only one) of these classes. Let us give one example: those whose nilradical is $\mathfrak{g}_{5.4}$. These algebras are listed in Tables 28 and 29 of \cite{Bock}. We have worked out some of their properties as before, and summarize them in Table \ref{tab:algsolv}. Let us also mention a typo in $\mathfrak{g}_{6.83}^{0,l}$, whose correct structure constants are provided in the following for completeness
\beq
\mathfrak{g}_{6.83}^{0,l}: \quad f^1{}_{24}=f^1{}_{35}=1 \ ,\ f^2{}_{26}=f^3{}_{36}=-f^4{}_{46}=-f^5{}_{56}= l \ , \ f^3{}_{26}=-f^4{}_{56}=1 \ .
\eeq
In this algebra, as well as others, appears a parameter $l$. We comment in Appendix \ref{ap:par} on various subtleties related to such parameters.

\begin{table}[H]
  \begin{center}
    \begin{tabular}{|c|c|c|c|c|}
    \hline
Algebra & $\mmm$ compactness & CS & DS & Eigenvalue \\
\hhline{=====}
$\mathfrak{g}_{6.83}^{0,l}$ ($l\neq0$) & ? & 6,5 & 6,5,1,0 & $4 l^2$ \\
\hline
$\mathfrak{g}_{6.84}$ & ? & 6,4,3 & 6,4,1,0 & $2$ \\
\hline
$\mathfrak{g}_{6.88}^{0,\mu_0,\nu_0}$ & $\mu_0 \nu_0 \neq 0$ & 6,5 & 6,5,1,0 & $4(\mu_0^2 - \nu_0^2)$ \\
($|\mu_0| + |\nu_0| \neq 0$) & & & & \\
\hline
$\mathfrak{g}_{6.89}^{0,\nu_0,s}$ & $\nu_0 \neq 0$ & $s\nu_0 \neq 0$: 6,5 & $s\nu_0 \neq 0$: 6,5,1,0 & $2(s^2 - \nu_0^2)$ \\
($|s| + |\nu_0| \neq 0$) & & $s\nu_0 = 0$: 6,3 & $s\nu_0 = 0$: 6,3,1,0 & \\
\hline
$\mathfrak{g}_{6.90}^{0,\nu_0}$ & $\nu_0 \neq 0$ & $\nu_0 \neq 0$: 6,5 & $\nu_0 \neq 0$: 6,5,1,0 & $2(1 - \nu_0^2)$ \\
 & & $\nu_0 = 0$: 6,3 & $\nu_0 = 0$: 6,3,1,0 & \\
\hline
$\mathfrak{g}_{6.91}$ & ? & 6,5 & 6,5,1,0 &  \\
\hline
$\mathfrak{g}_{6.92}^{0,\mu_0,\nu_0}$ ($\mu_0\nu_0 \neq 0$) & $\checkmark$ & 6,5 & 6,5,1,0 & $-4 \mu_0 \nu_0$ \\
\hline
$\mathfrak{g}_{6.92^*}^{0}$ & $\checkmark$ & 6,5 & 6,5,1,0 & $-4$ \\
\hline
$\mathfrak{g}_{6.93}^{0,\nu_0}$ & $|\nu_0| > \frac{1}{2}$ & $\nu_0 \neq 0$: 6,5 & $\nu_0 \neq 0$: 6,5,1,0 & $2(1 - 2\nu_0^2)$ \\
 & & $\nu_0 = 0$: 6,3 & $\nu_0 = 0$: 6,3,1,0 & \\
   \hline
    \end{tabular}
     \caption{6-dimensional real unimodular (indecomposable) solvable Lie algebras with nilradical $\mathfrak{g}_{5.4}$, from Tables 28 and 29 of \cite{Bock}, and their properties. Conditions imposed on the parameters are further discussed in Appendix \ref{ap:par}. The compactness of an associated group manifold $\mmm$, when settled, is read from the last remarks of Section 8.3 and 8.4 in \cite{Bock}; the conditions indicated here on parameters are sufficient, we do not know if they are necessary. We also give the CS and DS. The Killing form for these algebras admits at most one non-zero eigenvalue, that we indicate explicitly.}\label{tab:algsolv}
  \end{center}
\end{table}

\subsection{Method and tools for the identification}\label{sec:mettools}

\subsubsection{General method}\label{sec:metgen}

A 6d Lie algebra obtained in one of our solutions is given in terms of its structure constants $f^a{}_{bc}$ in an arbitrary basis of vectors $\left\{ E_a \right\}$. It must correspond to one (and only one) of the algebras appearing in the tables of \cite{Bock,mathbook}. Those are however given a priori in a different basis $\left\{ E'_a \right\}$, typically with a minimal amount of non-zero structure constants ${f^a{}_{bc}}'$. This is the reason why the identification of our algebras is challenging. The two algebras are isomorphic if and only if there exists a change of basis $M$ such that
\begin{equation}
E'_a = E_b \, (M^{-1})^b{}_a  \ \Leftrightarrow \ e^{a'}= M^a{}_b \, e^b  \,,  \quad \det M \neq 0 \,,
\end{equation}
where we also indicate the transformation of 1-forms $\left\{ e^a \right\}$. The two sets of structure constants are equivalently related in the following way
\begin{equation}
f^{a}{}_{de} = (M^{-1})^a{}_k M^b{}_d \, M^c{}_e\, {f^k{}_{bc}}'  \,. \label{changebasis}
\end{equation}
The relations \eqref{changebasis} amount to a high number of non-linear equations (depending on $M$), and it would be computationally too involved to try to solve them for every tabulated algebra. Rather, the method will consist in making use of the (basis) invariants and ideals defined in Section \ref{sec:algrecap} to reduce as much as possible the number of candidate algebras among the tabulated ones. Only then, and if there is more than one candidate algebra, we will find an explicit change of basis verifying \eqref{changebasis}.

\begin{figure}[H]
\hspace{-0.5in}	\centering{
		\resizebox{160mm}{!}{\fontsize{9pt}{11pt}\selectfont
			\def\svgwidth{5in}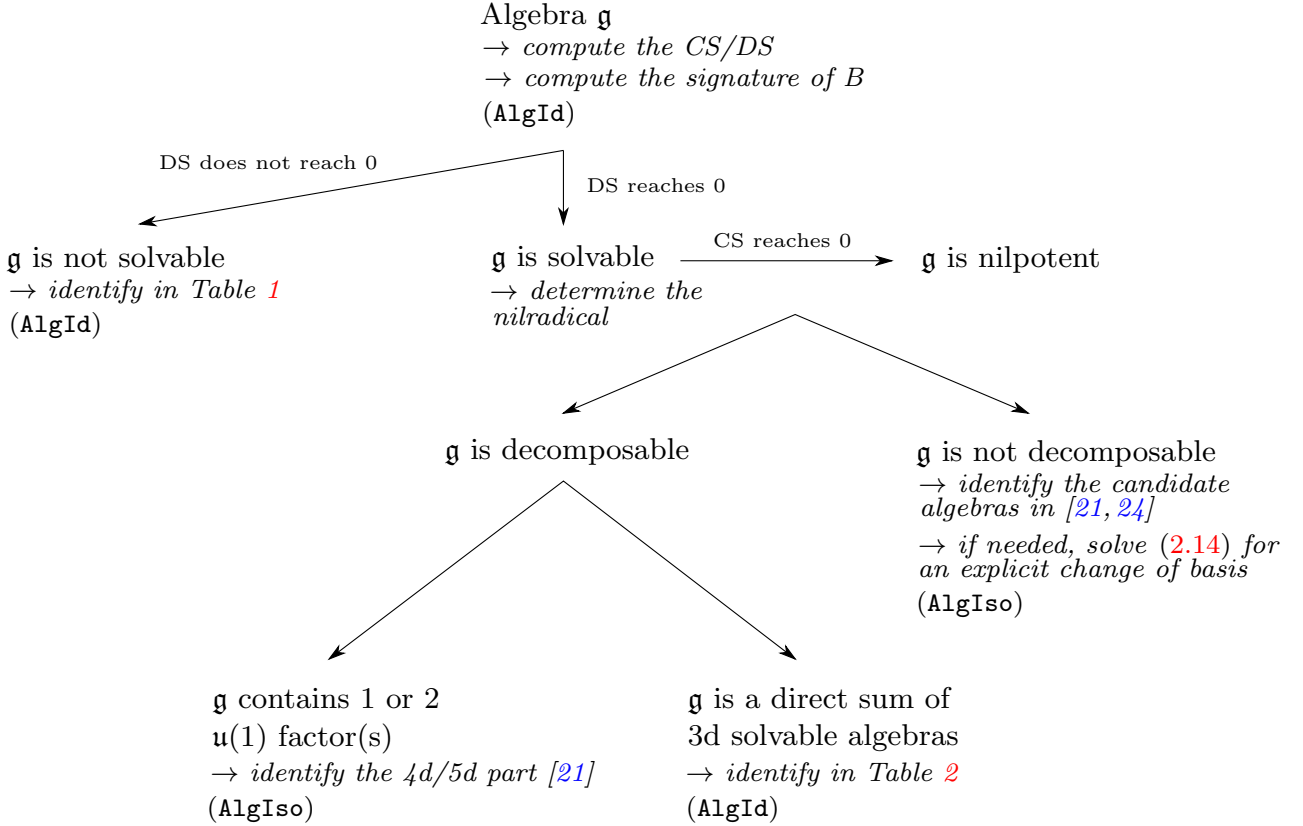}
		\caption{Summary of the procedure used for the identification of 6d unimodular Lie algebras.}\label{fig:method}
	}
\end{figure}

In more detail, to identify a 6d Lie algebra obtained in one solution in terms of its structure constants, we proceed as follows:
\begin{itemize}
\item We start by computing its CS, DS and the eigenvalues of its Killing form. The DS tells us if it is solvable or not. If it is not, it must be one of the 16 listed in Table \ref{tab:alg}. As can be seen there, the properties of the algebras allow to discriminate among all of them except for two. In this last case, we need an explicit change of basis to conclude. Apart from the latter, the procedure described so far has been automatized into the numerical tool {\tt AlgId}, that we present below.

\item In case the 6d algebra is solvable, we identify its nilradical. This is easily done using the definition given in Section \ref{sec:algrecap} and comparing to tables of nilpotent algebras. If necessary, one can compute as well the CS, DS of the nilradical to help in this comparison. If the CS indicates that the 6d algebra is nilpotent, then the nilradical is the algebra itself. Let us now consider that we face a solvable non-nilpotent algebra. In case it is indecomposable, one uses the nilradical to find a table of candidate algebras in \cite{Bock,mathbook}. For all these candidate algebras, the CS, DS and eigenvalues can again be computed and compared: see for instance Table \ref{tab:algsolv}. Once one is left with only few candidate algebras, an explicit change of basis needs to be found towards one of them. A dedicated numerical tool, {\tt AlgIso}, is presented below, providing a numerical matrix $M$ verifying \eqref{changebasis}. Another option is to find an analytical change of basis; we will give a few examples below.

\item A last possibility is that the 6d (real unimodular) non-nilpotent solvable algebra is decomposable. In that case, the tables of 6d indecomposable algebras are not useful and one has to devise what are the possibilities, according to the nilradical. One particular case is that of a decomposition into two 3d solvable algebras: such algebras are listed in Table \ref{tab:algsolv3d}, and their comparison is automatized in the tool {\tt AlgId}.
\end{itemize}
We summarize our procedure in Figure \ref{fig:method}.

\subsubsection{Numerical tools {\tt AlgId} and {\tt AlgIso}}\label{sec:numAlg}

We present in the following the numerical tools {\tt AlgId} and {\tt AlgIso} that we developed to help us identifying the algebras of our solutions. These tools are however built for a broader use.

The main input of the code {\tt AlgId} is a set of structure constants of a Lie algebra, as well as its dimension. The first part of the code computes the CS, DS, the Killing form eigenvalues, and its signature. In a second part, these results are compared to a data base of 6d algebras, namely the 16 of Table \ref{tab:alg} and the 9 of Table \ref{tab:algsolv3d} (without $6\, \mathfrak{u}(1)$), and the code indicates any match. These outputs should help identifying the initial algebra.

This code is meant to work not only for structure constants taking integer or round values, as e.g.~in classification tables, but also when having numerical values as those obtained in our solutions. Because of the latter, a precision parameter $\epsilon$ needs to be specified, and the code then interprets as vanishing, or sets to zero, various key quantities smaller than $\epsilon$: this avoids problems related to ``numerical zeros'' which are not exactly zero. In our case, a suitable value turns out to be $\epsilon = 10^{-10}$.

To compute the CS and DS of an algebra $\g$ with vectors $\left\{ E_a \right\}$, the code proceeds as follows. The CS and DS are the set of dimensions of the series of ideals $\g_{(k)}$ and $\g^{(k)}$. Each of these ideals is obtained by brackets between elements of subalgebras of $\g$. The idea of the code is to compute all these brackets, and store their result. More precisely, each bracket gives a vector, which is stored as the rows of a matrix $M_v$, when expressed in the initial basis $\left\{ E_a \right\}$. Here is an example
\beq
\begin{array}{l}
\left[E_1, E_2\right] = f^3{}_{12} E_3 \\
\left[E_1, E_3\right] = f^2{}_{13} E_2 + f^4{}_{13} E_4 \\
\quad \vdots
\end{array}
\quad
\rightarrow
\quad
M_v = \left( \begin{array}{ccccc}
0 & 0 & f^3{}_{12} & 0 & \dots \\
0 & f^2{}_{13} & 0 & f^4{}_{13} & \dots \\
 & & \vdots & &
\end{array} \right) \ .
\eeq
Determining the dimension of one ideal amounts to finding how many of these vectors are linearly independent. This could be done by computing the rank of the matrix $M_v$. But this option is not flexible enough in the case of numerical input, for instance in the case where two vectors are very similar, and should be equal up to a numerical error. What is rather done is then to complete the matrix $M_v$ into a square one, by determining the orthogonal space to the rows of $M_v$, and then compute the determinant of the square matrix. If the determinant is too small, the code considers it to be zero and decides that the vectors in $M_v$ are not independent. On the way, the vectors are normalised to 1 to allow for a fair evaluation of the determinant. This method to determine the linear independence is actually implemented every time a new vector is added to $M_v$: if it is found linearly independent, the dimension of the ideal is increased by one; if the vector is not independent, then it is not added to $M_v$. This is done recursively until one has tested all vectors obtained by all brackets defining the ideal. One obtains in this manner the dimension of the ideal, and builds this way the CS and DS of the algebra considered.

The rest of {\tt AlgId} is straightforward, so let us now present {\tt AlgIso}. That code aims at finding an isomorphism between two algebras that are specified as input. If one of the algebra depends on parameters, as e.g.~some of the tabulated ones, these parameters are automatically turned into variables and the code also searches for appropriate values. To find an isomorphism between the two algebras, the method amounts to solving the equations \eqref{changebasis}, where the variables are the matrix elements of $M$, the change of basis. To solve these equations, we proceed via a two-step minimisation of a loss function, built from the equations to solve. More details on this procedure can be found e.g.~in \cite{Andriot:2022way}, where this approach was used to find supergravity solutions. If a solution to the equations is found, then the isomorphism $M$ and possible values of algebra parameters are provided.

\subsubsection{Analytical changes of basis}

While most non-nilpotent solvable algebras are identified, following the method described in Section \ref{sec:metgen}, thanks to a final numerical change of basis, a few can still be determined by an analytical one. We present some examples below. In Appendix \ref{ap:chgebasis}, we present another one for a non-solvable algebra. For the latter, Table \ref{tab:alg} is enough for the identification. The analytical change of basis in that case can still serve further purposes, such as a classicality study \cite{Andriot:2020vlg}.

\subsubsection*{Solutions $s^+_{6666}3, 4$}

We start with solutions $s^+_{6666}3, 4$ which have the following non-zero structure constants
\beq
\label{scs6666}
s^+_{6666}3, 4: \quad {f^1}_{45},{f^5}_{14},{f^3}_{46},{f^6}_{34},{f^2}_{13},{f^2}_{56} \, .
\eeq
From this set, one identifies the nilradical to be $\mathfrak{g}_{5.4}$, with directions and non-zero structure constants
\beq
\mathfrak{n}=\{1,2,3,5,6\}, \quad {f^2}_{13},{f^2}_{56}\, .
\eeq
These algebras are thus among those of Table \ref{tab:algsolv}. To completely identify them, we determine in the following an analytical isomorphism. A first step is a relabeling on the set \eqref{scs6666}
\beq
1 \to 2, 2 \to 1, 3 \to 4, 4 \to 6, 5 \to 3, 6 \to 5: \quad {f^2}_{63},{f^3}_{26},{f^4}_{65},{f^5}_{46},{f^1}_{24},{f^1}_{35} \, .
\eeq
This small set of structure constants obeys a few relations, thanks to the Jacobi identities
\bea
& f^1{}_{e[3}f^e{}_{46]} = 0\  \Leftrightarrow\ \frac{f^1{}_{53}}{f^1{}_{24}} = \frac{f^2{}_{36}}{f^5{}_{46}} \ , \\
& f^1{}_{e[2}f^e{}_{56]} = 0\  \Leftrightarrow\ \frac{f^1{}_{53}}{f^1{}_{24}} = \frac{f^4{}_{56}}{f^3{}_{26}} \ .
\eea
We then perform the following rescaling
\beq
e^{a\neq 1',4'}=e^a\ ,\ e^{1'}=\frac{1}{{f^1}_{35}}\, e^1\ ,\ e^{4'}=\frac{f^1{}_{24}}{f^1{}_{53}}\, e^4\ ,
\eeq
leading, thanks to the above relations, to the new structure constants
\beq
{{f^2}_{36}}'= - {{f^5}_{46}}' = {f^2}_{36} \ ,\ \ {{f^3}_{26}}' = -{{f^4}_{56}}'= {f^3}_{26}\ ,\ \ {{f^1}_{35}}'= {{f^1}_{24}}'=1\ .
\eeq
We introduce the parameters $\mu_0=-{f^2}_{36}$ and $\nu_0={f^3}_{26}$. Given that $\mu_0 \neq \nu_0$ in our solutions, we identify the algebra to be $\mathfrak{g}_{6.92}^{0, \mu_0, \nu_0}$, for both solutions.

\subsubsection*{Solution $m_{5577}^{+\, *} 1$}

For this solution, we have the following set of non-vanishing structure constants:
\begin{equation}
m_{5577}^{+\, *} 1: \qquad \f{6}{23} = \f{6}{14} \,, \,\f{2}{35} =\f{1}{45} \,, \,\f{4}{15} = \f{3}{25} \,.
\end{equation}
The nilradical is identified to be $\mathfrak{g}_{5.4}$ with
\begin{equation}
\mathfrak{n} = \left\{1,2,3,4,6  \right\} \,, \quad \text{and} \quad \f{6}{23} = \f{6}{14} \,,
\end{equation}
so the algebra can be found in Table \ref{tab:algsolv}. One can perform the following relabeling of directions
\begin{equation}
\hspace{-0.2in} 1 \rightarrow 2\,,2 \rightarrow 3\,,3 \rightarrow 5\,,4 \rightarrow 4\,,5 \rightarrow 6\,,6 \rightarrow 1 :\quad \f{1}{35} = \f{1}{24} \,, \,\f{3}{56} =\f{2}{46} \,, \,\f{4}{26} = \f{5}{36} \,.
\end{equation}
We have in addition the following signs: $\f{3}{56} >0, \f{5}{36} < 0$. We then perform the following rescaling on forms
\beq
{e^1}' = \frac{1}{\f{1}{35} \sqrt{-\f{3}{56} \f{5}{36}}}\, e^1 \ ,\ {e^{2,3}}' = \frac{1}{\sqrt{\f{3}{56}}} e^{2,3} \ ,\ {e^{4,5}}' = \frac{1}{\sqrt{- \f{5}{36}}}\, e^{4,5} \ ,\ {e^6}' = \sqrt{-\f{3}{56} \f{5}{36}}\, e^6 \ .
\eeq
The new normalization allows to directly identify the algebra to be $\mathfrak{g}^0_{6.92^*}$.

\subsection{Results}\label{sec:resultsalg}

Using the method and tools described in Section \ref{sec:metgen}, we have identified all algebras of the solutions found in \cite{Andriot:2022way}, as well as the algebras of the previously found solutions $s^0_{55}$ 1 and $s^+_{55}$ 1 - 28 \cite{Andriot:2020wpp,Andriot:2021rdy}. This allows in particular to discuss the compactness of the group manifolds, using the material of Section \ref{sec:algex} or further results in \cite{Bock}. We summarize our findings as follows.

\subsubsection*{De Sitter solutions}

\begin{table}[H]
  \begin{center}
    \begin{tabular}{|c||c|c|c|c|c|}
    \hline
algebra & $\mathfrak{so}(3) \oplus \mathfrak{so}(3)$ & $\mathfrak{so}(3) \oplus 3\, \mathfrak{u}(1)$ & $\mathfrak{so}(3) \oplus {\rm Heis}_3$ & $\mathfrak{so}(3) \oplus \mathfrak{g}_{3.4}^{-1}$ & $\mathfrak{g}_{3.5}^{0} \oplus \mathfrak{g}_{3.5}^{0}$ \\
\hhline{=::=====}
solutions & $s^+_{6666}$ 1 & $m^+_{46}$ 10 & $s^+_{55}$ 20, 21 & $s^+_{55}$ 19 & $s^+_{55}$ 14 \\
\hhline{=::=====}
compactness & $\checkmark$ & $\checkmark$ & $\checkmark$ & $\checkmark$ & $\checkmark$ \\
    \hline
  \multicolumn{6}{c}{}\\
    \hline
algebra & $\mathfrak{g}_{3.4}^{-1} \oplus \mathfrak{g}_{3.4}^{-1}$ & $\mathfrak{g}_{3.4}^{-1} \oplus \mathfrak{g}_{3.5}^{0}$ & $\mathfrak{g}_{6.76}^{-1}$ & $\mathfrak{g}_{6.92}^{0,\mu_0,\nu_0}$  & $\mathfrak{g}_{6.92^*}^0$  \\
\hhline{=::=====}
solutions &  $s^+_{55}$ 15 & $s^+_{55}$ 22 - 27 & $s^+_{55}$ 16, 17 & $s^+_{6666}$ 3, 4, $m^+_{5577}$ 3 - 6 & $m_{5577}^{+\,*}$ 1 \\
\hhline{=::=====}
compactness & $\checkmark$ & $\checkmark$ & $\checkmark$ & $\checkmark$ & $\checkmark$ \\
    \hline
    \end{tabular}
     \caption{Algebras identified in de Sitter solutions, leading to compact group manifolds.}\label{tab:comp+}
  \end{center}
\end{table}

\begin{table}[H]
  \begin{center}
    \begin{tabular}{|c||c|c|c|c|}
    \hline
algebra & $\mathfrak{so}(3,1)$ & $\mathfrak{so}(2,1) \oplus \mathfrak{so}(2,1)$ & $\mathfrak{so}(3) \oplus \mathfrak{so}(2,1)$ & $\mathfrak{so}(2,1) \oplus 3\mathfrak{u}(1)$ \\
\hhline{=::====}
solutions & $m^+_{55}$ 1, $s^+_{6666}$ 2 & $m^+_{55}$ 2 - 4, $m^+_{5577}$ 2, 7, 12 & $m^+_{5577}$ 1 & $m^+_{46}$ 1 - 9 \\
\hhline{=::====}
compactness & $\times$ & $\times$ & $\times$ & $\times$  \\
    \hline
  \multicolumn{5}{c}{}\\
    \hhline{----~}
    algebra & $\mathfrak{so}(2,1) \oplus {\rm Heis}_3$ & $\mathfrak{so}(2,1) \oplus \mathfrak{g}_{3.5}^{0}$ & $\mathfrak{so}(2,1) \oplus \mathfrak{g}_{3.4}^{-1}$ & \multicolumn{1}{c}{} \\
\hhline{=::===~}
solutions & $s^+_{55}$ 18 & $s^+_{55}$ 12, $m^+_{5577}$ 9, 10 & $s^+_{55}$ 1 - 11, 13, 28, & \multicolumn{1}{c}{} \\
& &  & $m^+_{5577}$ 8, 11 & \multicolumn{1}{c}{} \\
\hhline{=::===~}
compactness & $\times$ & $\times$ & $\times$ & \multicolumn{1}{c}{} \\
    \hhline{----~}
    \end{tabular}
     \caption{Algebras identified in de Sitter solutions, leading to non-compact group manifolds.}\label{tab:noncomp+}
  \end{center}
\end{table}

Finally, the solution $s^+_{66}$ 1 was identified to be on $\mathfrak{g}_{6.88}^{0,\mu_0,0} = \mathfrak{g}_{6.88}^{0,1,0}$ (see Appendix \ref{ap:par}), but we do not know whether this algebra can provide compact group manifolds.

\subsubsection*{Minkowski solutions}

\begin{table}[H]
  \begin{center}
    \begin{tabular}{|c||c|c|c|c||c|c|}
    \hline
algebra & ${\rm Heis}_3 \oplus {\rm Heis}_3$ & $\mathfrak{g}_{6.88}^{0,\mu_0,\nu_0}$ & $\mathfrak{g}_{6.89}^{0, \nu_0,s}$ & $\mathfrak{g}_{5.14}^0 \oplus \mathfrak{u}(1)$ & $\mathfrak{so}(3)\oplus \mathfrak{so}(2,1)$ & $\mathfrak{so}(2,1)\oplus 3 \mathfrak{u}(1)$ \\
\hhline{=::====::==}
solutions & $s_{55}^0$ 1 & $s_{555}^0$ 4 & $s_{555}^0$ 2, 3 & $m_{466}^0$ 4 & $s_{555}^0$ 1 & $m_{46}^0$ 1 \\
\hhline{=::====::==}
compactness & $\checkmark$ & $\checkmark$ & $\checkmark$ & $\checkmark$ & $\times$ & $\times$  \\
    \hline
    \end{tabular}
     \caption{Algebras identified in Minkowski solutions, leading to compact ($\checkmark$) or non-compact ($\times$) group manifolds.}\label{tab:0alg}
  \end{center}
\end{table}

Several other Minkowski solutions were found with algebras that may or may not provide compact group manifolds: we refer in the following to related propositions in \cite{Bock} that could help in settling this matter, in case there is a particular interest in a specific solution. This is the situation encountered for solutions $m_{46}^0$ 2 and $m_{466}^0$ 3, 5 with algebra $\mathfrak{g}_{5.17}^{0, 0, r} \oplus \mathfrak{u}(1)$ (Prop. 7.2.13), $m_{466}^0$ 1, 2 with $\mathfrak{g}_{5.13}^{-1, 0, r} \oplus \mathfrak{u}(1)$
(Prop. 7.2.6), and $m_{466}^0$ 6 with $\mathfrak{g}_{5.7}^{p, -p, -1} \oplus \mathfrak{u}(1)$ (Prop. 7.2.1).

\subsubsection*{Anti-de Sitter solutions}

\begin{table}[H]
  \begin{center}
    \begin{tabular}{|c||c|c|c|c|}
    \hline
algebra & $\mathfrak{so}(3)\oplus 3 \mathfrak{u}(1)$ & $\mathfrak{so}(3)\oplus \mathfrak{g}_{3.5}^0$ & $\mathfrak{g}_{3.5}^0 \oplus \mathfrak{g}_{3.5}^0$ & $\mathfrak{g}_{6.10}^{0,0}$ \\
\hhline{=::====}
solutions & $m_{46}^-$ 1, 2 & $s_{55}^-$ 1 & $s_{55}^-$ 2, 3, 4 & $m_{46}^-$ 4, 5 \\
\hhline{=::====}
compactness & $\checkmark$ & $\checkmark$ & $\checkmark$ & $\checkmark$ \\
    \hline
    \end{tabular}
     \caption{Algebras identified in anti-de Sitter solutions, leading to compact group manifolds.}\label{tab:comp-}
  \end{center}
\end{table}

Finally, solution $m_{46}^-$ 3 was found on the algebra $\mathfrak{g}_{6.34}^{0,0,0}$, and we do not know if the latter can provide a compact manifold.

\section{Stability}\label{sec:stab}

In this section, we present the key elements of the 4d effective action \eqref{S4dgen}, obtained after dimensional reduction and consistent truncation of our 10d solutions, allowing us to determine their stability. We first discuss in Section \ref{sec:pot} the scalar fields considered and the scalar potential $V$, then in Section \ref{sec:met} the field space metric $g_{ij}$ and the problem of field redundancy. We present in Section \ref{sec:numMSSSp} the numerical tool that we have developed for these computations, and we finally discuss in Section \ref{sec:resultstab} the stability of our solutions, inferred from these considerations, and compare it to various conjectures.

\subsection{Scalar fields and potential}\label{sec:pot}

As explained in the Introduction, we consider a restricted set of 4d scalar fields $(\rho,\tau,\sigma_I)$, where $I=1,...,N$ runs over the sets of (parallel) sources. The 6d volume $\rho$ and 4d dilaton $\tau$ were introduced together with their potential in \cite{Hertzberg:2007wc}. The $\sigma_I$, related to internal volumes wrapped by the sources, can be defined independently: they were introduced and motivated in \cite{Danielsson:2012et, Junghans:2016uvg}. The generic scalar potential depending on $\sigma_I$ was obtained in \cite{Andriot:2018ept} for $N=1$, and in \cite{Andriot:2019wrs} for $N>1$, with a single source dimensionality $p$; here we will extend it to multiple dimensionalities.

All these scalar fields are obtained as specific fluctuations around the background 6d metric and dilaton. To obtain the 4d scalar potential, one should introduce these fluctuations in the 10d action. A first result is the following potential $V$ depending on $\rho,\tau$
\beq
\frac{2}{M_p^2} V =  - \tau^{-2} \bigg( \rho^{-1} {\cal R}_6 -\frac{1}{2} \rho^{-3} |H|^2 \bigg) - g_s \tau^{-3} \sum_{p,I} \rho^{\frac{p-6}{2}} \frac{T_{10}^{(p)_I}}{p+1} +\frac{1}{2} g_s^2 \tau^{-4} \sum_{q=0}^{6} \rho^{3-q} |F_q|^2  \label{pot1} \ ,
\eeq
while ${\cal R}_6, H, F_q, T_{10}^{(p)_I}$ should still be fluctuated with respect to the $\sigma_I$; we will do so in the following. Let us emphasize that in this potential, the terms in $F_5$ or $F_6$ are not obtained in the same way as the others, because of the contribution of corresponding 4d components; we refer the interested reader to the appendix of \cite{Andriot:2020lea}, which completes the derivation of the potential. It is shown there that eventually, these terms can be recast in the same form as the other $F_q$, including the fluctuation to come with respect to $\sigma_I$, so we treat here all fluxes together.

Each $\sigma_I$ is defined with respect to a given set $I$ of parallel sources. For this reason, while fluctuations with respect to $\sigma_I$ were generically described in the aforementioned papers, the resulting potential (and kinetic terms described in Section \ref{sec:met}) is dependent on each specific source configuration. We need here these results for each solution class of \cite{Andriot:2022way}, because we have found solutions on a compact manifold for almost all of them (see Section \ref{sec:resultsalg}), justifying the study of a corresponding 4d theory. In addition, the formulas of \cite{Andriot:2019wrs} need a slight generalization to the case of multiple dimensionalities, as we encounter in some solution classes. For these reasons, we present here once again the definition of these fields and corresponding fluctuations, introducing however new notations and tools allowing a more systematic treatment for any solution class. This will be used in the numerical tool {\tt MaxSymSolSpec} ({\tt MSSSp}) that we have developed, to provide the potential for any source configuration.

For each set $I$ of $p$-sources, with certain parallel and transverse directions, one defines as follows a 4d scalar fluctuation $\sigma_I$ on the 6d vielbeins
\beq
e^{a_{||_I}}{}_m \rightarrow {\sigma_I}^\frac{A_I}{2}\ e^{a_{||_I}}{}_m \ ,\quad e^{a_{\bot_I}}{}_m \rightarrow {\sigma_I}^\frac{B_I}{2}\ e^{a_{\bot_I}}{}_m \ ,\quad A_I= p-9 \ , \ B_I=p-3 \ .
\eeq
The exponents $A_I$ and $B_I$ are chosen in such a way that the determinant $|g_6|$ is left invariant under this fluctuation. This should be done for all sets $I$ of sources (with possibly different $p$). Overall, each 6d vielbein $e^{a}{}_m$ gets multiplied by a product of powers of $\sigma_I$ that we denote $\pi_{a}$, as follows
\beq
e^{a}{}_m \rightarrow \pi_{a}\ e^{a}{}_m \ \mbox{(no sum)} \ ,\ {\rm where}\ \pi_{a} = \prod_I {\sigma_I}^{\frac{P_I(a)}{2}} \ ,\ P_I(a_{||_I}) = A_I \ ,\  P_I(a_{\bot_I}) = B_I \ . \label{pia}
\eeq
Introducing these $\pi_{a}$ is a convenient novelty. From there, one gets the fluctuations of each quantity entering the potential \eqref{pot1} by going to the orthonormal coframe and following the vielbein dependence:
\beq
H_{abc} \rightarrow (\pi_{a} \pi_{b} \pi_{c} )^{-1}\  H_{abc} \ ,\ F_{q\, a_1 \dots a_q} \rightarrow (\pi_{a_1} \dots \pi_{a_q} )^{-1}\  F_{q\, a_1 \dots a_q} \ , \ f^a{}_{bc} \rightarrow \pi_{a} (\pi_{b} \pi_{c} )^{-1}\ f^a{}_{bc}
\eeq
The dependence in ${\cal R}_6$ is then obtained using the standard formula
\begin{equation}
\label{R6}
-2\, {\cal R}_6 = \delta^{ce} \, f^{b}{}_{ac} \, f^{a}{}_{be}   + \frac{1}{2} \, \delta^{eb} \,\delta^{fc} \, \delta_{ga} \, f^{g}{}_{ef} \, f^{a}{}_{bc}  \ ,
\end{equation}
while the square of fluxes in the potential give rise to the sum of the squares of fluctuated components. Finally, the fluctuation of the source term $T_{10}^{(p)_I}$ corresponds to that of the internal volume form ${\rm vol}_{||_I}$: we get for each set $I$
\beq
T_{10}^{(p)_I} \rightarrow  T_{10}^{(p)_I}\ \prod_{a= a_{||_I}} \pi_{a} \ .
\eeq
One deduces from these fluctuations and \eqref{pot1} the complete potential $V(\rho,\tau,\sigma_I)$, for each source configuration.

As an example, the complete potential for the solution class $s_{55}$ was given in \cite{Andriot:2021rdy}. Let us give here the potential for the class $m_{46}$ with only 1 $D_6$: interestingly, it admits sources of multiple dimensionalities. The sets are $I=1$ with an $O_4$ along 4 and contribution $T_{10}^{(4)}$, $I=2$ with an $O_6$ along 123 and $T_{10}^{(6)_1}$, $I=3$ with $D_6$ along 156 and $T_{10}^{(6)_2}$. The potential is then given by
\bea
\frac{2}{M_p^{2}} \, V(\rho, \tau, \sigma_1, \sigma_2, \sigma_3) = &-\tau^{-2}  \rho^{-1} {\cal R}_6(\sigma_1, \sigma_2, \sigma_3) \label{potential}\\
& +\frac{1}{2}\, \tau^{-2} \rho^{-3} \sigma_1^{-3} \sigma_2^{3} \left( \sigma_3^{-3}  \, |H^{(1)_3}|^2 + \sigma_3^{3}  \, |H^{(2)_3}|^2  \right) \nn\\
&- \, g_s \, \tau^{-3} \, \left(\rho^{-1} \, \sigma_1^{-\frac{5}{2}} \sigma_2^{\frac{3}{2}}  \sigma_3^{\frac{3}{2}}  \, \frac{T_{10}^{(4)}}{5}+  \sigma_1^{\frac{3}{2}} \sigma_2^{-\frac{9}{2}}  \sigma_3^{\frac{3}{2}}  \, \frac{T_{10}^{(6)_1}}{7} + \sigma_1^{\frac{3}{2}} \sigma_2^{\frac{3}{2}}  \sigma_3^{-\frac{9}{2}}  \, \frac{T_{10}^{(6)_2}}{7}  \right) \nn\\
&+ \frac{1}{2}g_s^2\, \tau^{-4} \left(\rho \sigma_1^{-2} \left( |F_2^{(1)_3}|^2 + \sigma_3^{6}  \, |F_2^{(2)_3}|^2  \right) + \rho^{-1} \sigma_1^{2} \left( \sigma_3^{-6} \, |F_4^{(1)_3}|^2 +  |F_4^{(2)_3}|^2  \right) \right) \ ,\nn
\eea
where we recall the flux notation $H^{(n)_{I=3}}$, of components having $n$ indices along the set $I=3$. We refer to the result given by ${\tt MSSSp}$ for the precise expression of ${\cal R}_6(\sigma_1, \sigma_2, \sigma_3)$ as a sum of powers of $\sigma_I$ times structure constants.

Let us finally recall that in our conventions, the background (i.e.~our solutions) is recovered at $\rho=\tau=\sigma_I=1$. Since we have a consistent truncation, this corresponds in 4d to a critical point $\del_{\phi^i} V =0$, while at this point one also has $\frac{2}{M_p^{2}} V = \frac{1}{2} {\cal R}_4 $. This is checked on each of our solutions.

\subsection{Field space metric and redundancy}\label{sec:met}

Following Appendix D of \cite{Andriot:2020wpp}, the kinetic terms appearing in \eqref{S4dgen} are given by
\beq
\frac{1}{M_p^2}\, g_{ij} \del_{\mu}\phi^i \del^{\mu}\phi^j = 2 \tau^{-2} (\del \tau)^2 + \frac{3}{2} \rho^{-2} (\del \rho)^2 - \frac{1}{4}\, \del_{\mu} (m_{ab}) \del^{\mu} ((m^{-1})^{ab}) \ , \label{kin}
\eeq
where $m_{ab}$ is the diagonal 6d metric in orthonormal coframe fluctuated with $\sigma_I$; it has determinant 1. Using the convenient notation introduced in \eqref{pia}, we obtain
\beq
m_{ab} = \pi_{a}^2\ \delta_{ab} \quad \text{(without sum)} \label{mab} \ .
\eeq
By $(m^{-1})^{ab}$ we denote in \eqref{kin} the coefficients of the inverse of $m$. It is then straightforward to obtain the kinetic terms. In particular, the expression for $m_{ab}$ leads to many cross terms $\del_{\mu} \sigma_I  \del^{\mu} \sigma_J $, i.e.~non-diagonal elements of the field space metric $g_{ij}$.\\

An issue is however that the fields $\sigma_I$ are sometimes redundant. This can be understood as follows: each $\sigma_I$ is a metric fluctuation, in correspondence with an internal volume wrapped by a source set. The independence of the $\sigma_I$ can be seen as the independence of these volumes: for instance if one has $O_5$ along 12, 34 and $D_7$ along 1234, the volumes are not independent and there would be a redundancy in the $\sigma_I$. This depends entirely on the source configuration, and for each of them, we need to specify a set of independent $\sigma_I$.

The problem of the redundancy is equivalently seen through the field space metric $g_{ij}$: it has vanishing determinant if there are redundant fields. Indeed, redundant fields can be removed by a field redefinition. But removing some $\sigma_J$ would lead to vanishing field metric coefficients along the $\del^{\mu} \sigma_J$, hence a vanishing determinant. A set of independent fields must then be identified before computing the metric. A concrete way to determine redundant fields is to find a field redefinition that removes one or more $\sigma_J$ fields completely from the $\pi_a$ defined in \eqref{pia} (or equivalently sets these fields to 1). Since the $\pi_a$ are the building blocks for the potential and the field space metric (see \eqref{mab}), fields $\sigma_J$ removed from the $\pi_a$ will not appear anywhere and were certainly redundant.

A field redefinition to remove fields $\{\sigma_{X}\}$ from $\pi_a$ and keep $\{\sigma_M\}$ can be designed as follows; it is not the most general, but it will be enough for our purposes
\beq
\sigma_M \rightarrow \sigma_M\, \prod_{X}\, \sigma_X^{s_{XM}} \ ,\quad \sigma_X \rightarrow \sigma_X \ . \label{fieldredef}
\eeq
One verifies that $\sigma_{X}$ are removed from all $\pi_a$ if and only if one finds exponents $s_{XM}$ satisfying
\beq
\forall a,\, X,\qquad P_X(a) + \sum_M s_{XM} P_M(a) = 0 \ . \label{fieldredefcond}
\eeq
Let us consider a first particular solution: $s_{XM}= 1$, and there is a single field to remove, the last one, i.e.~$X=N$. The field redefinition \eqref{fieldredef} becomes
\beq
\sigma_{I\neq N} \rightarrow \sigma_{I\neq N}\, \sigma_N \ ,\quad \sigma_N \rightarrow \sigma_N \ . \label{fieldredefstand}
\eeq
This field redefinition was used already successfully in $s_{6666}$ \cite{Danielsson:2012et, Andriot:2019wrs} and $s_{55}$ \cite{Andriot:2020wpp}. One verifies indeed that the condition \eqref{fieldredefcond} holds, with $\sum_I P_I(a) = 2(A+B)$ for $p=6$ and $\sum_I P_I(a) = 2B+A$ for $p=5$, and both vanish. We verify that the same holds for $m_{5577}$ and $m_{5577}^*$, allowing there again to remove the last field, curing the redundancy.

Other cases require different solutions to \eqref{fieldredefcond} to remove differently redundant fields, for instance when $\exists\, a$ s.t. $\sum_I P_I(a) \neq 0$. This happens for $m_{46}$ with $O_4$ along 4, $O_6$ along 123, and $D_6$ along 156, 256, 356; these five sets defining $\sigma_{1,...,5}$ respectively. The field space metric determinant vanishes for five $\sigma_I$, but not for four. We find the appropriate field redefinition \eqref{fieldredef} to take the form
\beq
\sigma_1 \rightarrow \sigma_1\, \sigma_5^3 \ ,\ \sigma_2 \rightarrow \sigma_2\, \sigma_5^2 \ ,\ \sigma_{3,4} \rightarrow \sigma_{3,4}\, \sigma_5 \ ,\ \sigma_5 \rightarrow \sigma_5 \ ,
\eeq
removing $\sigma_5$ from the $\pi_a$.

Another case is that of $m_{55}$ with 7 sets in the following order: $O_5$ along 12, 34, $D_5$ along 56, $D_7$ along 2456, 2356, 1456, 1356. Solutions have been found with all or some of these sets turned on. We consider the corresponding 7 fields $\sigma_I$. Solving \eqref{fieldredefcond}, we find the following general field redefinition
\beq
\sigma_{1,2} \rightarrow \sigma_{1,2}\, \sigma_3\, \sigma_5\, \sigma_7 \ ,\ \sigma_4 \rightarrow \sigma_4\, \sigma_7 \ ,\ \sigma_6 \rightarrow \sigma_6\, \sigma_5\ ,\ \sigma_{3,5,7} \rightarrow \sigma_{3,5,7} \ . \label{fieldredefIIB}
\eeq
It allows to remove $\sigma_{3,5,7}$ from the $\pi_a$, in the case where all sources are present. In the case where $T_{10}^3=0$, one can still use \eqref{fieldredefIIB}, setting $\sigma_3=1$ and removing $\sigma_{5,7}$. Similarly, for $T_{10}^3=T_{10}^5=0$, one can use \eqref{fieldredefIIB} setting $\sigma_{3,5}=1$ and removing $\sigma_{7}$. All these cases amount in the end to setting $\sigma_{3,5,7}$ to $1$.\footnote{In the case where $T_{10}^3=T_{10}^5=T_{10}^6=0$, one can use \eqref{fieldredefIIB}, setting $\sigma_{3,5,6}=1$ and removing $\sigma_{7}$: this redefinition matches the more standard one \eqref{fieldredefstand}. We however do not encounter this case in our solutions.} \\

Once we know which fields $\sigma_I$ are redundant and should be removed (or equivalently set to 1), we are left with a set of independent fields, and correspondingly a non-degenerate field space metric. Let us give this data in one example, with the source sets considered and ordered, the corresponding independent scalar fields, and the field space metric expressed in that field basis:

\hspace{0.1in}

$\boldsymbol{m_{46}}$ (1 $D_6$): $O_4$ (4), $O_6$ (123), $D_6$ (156), or $\boldsymbol{m_{466}}$: $O_4$ (4), $O_6$ (123, 156).

Fields: $(\rho, \tau, \sigma_1, \sigma_2, \sigma_3)$
\begin{equation}
g_{ij} = M_p^2\,
\begin{pmatrix}
\mathlarger{\frac{3}{2 \rho^2}} & 0 & 0 &0 & 0  \\[10pt]
0 & \mathlarger{\frac{2}{\tau^2}} & 0 &0 &0 \\[10pt]
0 & 0 & \mathlarger{\frac{15}{2 \sigma_1^2}} & \mathlarger{-\frac{9}{2\sigma_1 \sigma_2}} & \mathlarger{-\frac{9}{2\sigma_1 \sigma_3}} \\[10pt]
0 & 0 &\mathlarger{-\frac{9}{2\sigma_1 \sigma_2}} & \mathlarger{\frac{27}{2\sigma_2^2}} & \mathlarger{-\frac{9}{2\sigma_2 \sigma_3}}  \\[10pt]
0 & 0 &\mathlarger{-\frac{9}{2\sigma_1 \sigma_3}} & \mathlarger{-\frac{9}{2\sigma_2 \sigma_3}} & \mathlarger{\frac{27}{2\sigma_3^2}}
\end{pmatrix} \ . \label{gijm461}
\end{equation}

\noindent The data for the other cases encountered in our solutions is given in Appendix \ref{ap:gij}; we refer to the code {\tt MSSSp} for further cases.

\subsection{Numerical tool {\tt MaxSymSolSpec} ({\tt MSSSp})}\label{sec:numMSSSp}

The computation of the scalar potential and the field space metric, as described in Section \ref{sec:pot} and \ref{sec:met}, has been automatized in the numerical tool {\tt MaxSymSolSpec} ({\tt MSSSp}) that we have developed. The code first takes as input the list of source sets. From this data, the fields $(\rho, \tau, \sigma_I)$ can be defined. A first task is to determine a set of independent fields, and remove the redundant ones. The user can specify a complete list of redundant fields, based for instance on Section \ref{sec:met} and Appendix \ref{ap:gij}. Otherwise, the code determines such a list by itself. To that end, the field space metric is computed and its rank is checked, row after row, allowing to identify redundant fields. Once a set of independent fields is identified, a proper field space metric is computed, as well as the scalar potential $V$. The latter is obtained by considering the fluctuations $\pi_a$ as described in Section \ref{sec:pot}.

With a set of independent fields, the corresponding field space metric and the scalar potential, the code can compute the mass spectrum, following definitions of Section \ref{sec:resultstab}. This is done for a 10d supergravity solution provided as an input. The code verifies that it is a critical point of the potential. It then computes the parameter $\eta_V$, the masses${}^2$ and their associated field space eigenvectors. Note that the mass matrix $M$ transforms covariantly under (field space) diffeomorphisms, i.e.~field redefinitions. So its eigenvalues, namely the mass spectrum, and in particular the value of $\eta_V$, are unchanged when choosing a different (diffeomorphic) set of independent fields.

\subsection{Results: stability of the solutions and (swampland) conjectures}\label{sec:resultstab}

Having determined the scalar potential and the field space metric of the 4d effective theory \eqref{S4dgen}, for a set of independent scalar fields $(\rho,\tau,\sigma_I)$, we can compute the corresponding mass spectrum for each solution of \cite{Andriot:2022way}. It is given by the eigenvalues (masses${}^2$) of the mass matrix, with coefficients $M^i{}_j = g^{ik} \nabla_{\!\phi^k} \del_{\phi^j} V$, at the critical point $\rho=\tau=\sigma_I=1$. The connection term due to $\nabla$ vanishes at an extremum, since it is proportional to a first derivative of the potential. Therefore, one only needs to compute the eigenvalues of $g^{-1}$ times the Hessian of the potential $V$, at this point. All these computations are performed using {\tt MSSSp}.

From the mass spectrum, one reads the stability of the solution (at least due to this set of scalar fields). For de Sitter and anti-de Sitter solutions where $V \neq 0$, this is better captured by the parameter $\eta_V$ that we recall here
\beq
\eta_V= {M_p}^2\, \frac{{\rm min}\ \nabla \del V}{V} \ , \label{etaV}
\eeq
where the numerator stands for the minimal eigenvalue among the masses${}^2$. The $\eta_V$ is computed at the critical point. Note that we use the same definition for de Sitter and anti-de Sitter extrema, although the sign of $V$ changes.

Let us finally recall from \cite{Andriot:2020wpp} that the minimal eigenvalue of a mass matrix can only get lowered if one adds more fields. Therefore, if an instability is detected within our set of fields, it will not be cured with more fields, and we can conclude on a unstable solution. We now study the stability of each solution of \cite{Andriot:2022way} according to the sign of the cosmological constant.

\subsubsection{De Sitter}

The values of $\eta_V$ for each de Sitter solution of \cite{Andriot:2022way} are given in Table \ref{tab:etaIIA+} and \ref{tab:etaIIB+}.

\begin{table}[H]
  \begin{center}
    \begin{tabular}{|c||c|c|c|c|c|c|c|c|c|}
    \hline
class & \multicolumn{1}{c||}{$s_{66}^+$} & \multicolumn{4}{c||}{$s_{6666}^+$} & \multicolumn{4}{c|}{$m_{46}^+$}  \\
\hhline{-||-||----||----}
solution & \multicolumn{1}{c||}{1} & 1 & 2 & 3 & \multicolumn{1}{c||}{4} & 1 & 2 & 3 & 4 \\
    \hhline{=::=::====::====}
$-\eta_V$ & \multicolumn{1}{c||}{3.6170} & 18.445 & 2.6435 & 2.3772 & \multicolumn{1}{c||}{3.6231} & 3.6764 & 3.7145 & 2.2769 & 2.8266 \\
    \hline
  \multicolumn{10}{c}{}\\
\hhline{-------~~~}
class & \multicolumn{6}{c|}{$m_{46}^+$} & \multicolumn{3}{c}{} \\
\hhline{-||------|~~~}
solution  & 5 & 6 & 7 & 8 & 9 & \multicolumn{1}{c|}{10} & \multicolumn{3}{c}{} \\
    \hhline{=::======:~~~}
$-\eta_V$ & 0.36462 & 3.0124 & 2.0672 & 2.3554 & 2.6418 & \multicolumn{1}{c|}{1.2539} & \multicolumn{3}{c}{} \\
\hhline{-------~~~}
    \end{tabular}
     \caption{Values of $-\eta_V$ obtained with the set of independent fields considered for each de Sitter solution in type IIA.}\label{tab:etaIIA+}
  \end{center}
\end{table}

\begin{table}[H]
  \begin{center}
    \begin{tabular}{|c||c|c|c|c|c|c|c|c|c|}
    \hline
class & \multicolumn{1}{c||}{$s_{55}^+$} & \multicolumn{4}{c||}{$m_{55}^+$} & \multicolumn{4}{c|}{$m_{5577}^+$} \\
\hhline{-||-||----||----}
solution & \multicolumn{1}{c||}{28} & 1 & 2 & 3 & \multicolumn{1}{c||}{4} & 1 & 2 & 3 & 4 \\
    \hhline{=::=::====::====}
$-\eta_V$ & \multicolumn{1}{c||}{3.2374} & 2.5435 & 2.6059 & 2.7126 & \multicolumn{1}{c||}{3.3574} & 4.7535 & 3.5034 & 3.2722 & 3.1779 \\
    \hline
  \multicolumn{10}{c}{}\\
    \hline
class & \multicolumn{8}{c||}{$m_{5577}^+$} & \multicolumn{1}{c|}{$m_{5577}^{*\,+}$} \\
\hhline{-||--------||-}
solution  & 5 & 6 & 7 & 8 & 9 & 10 & 11 & \multicolumn{1}{c||}{12} & \multicolumn{1}{c|}{1} \\
    \hhline{=::========::=}
$-\eta_V$  & 4.7957 & 4.9129 & 3.4210 & 3.5611 & 2.9333 & 2.9003 & 3.4806 & \multicolumn{1}{c||}{2.8966} & \multicolumn{1}{c|}{5.0483} \\
    \hline
    \end{tabular}
     \caption{Values of $-\eta_V$ obtained with the set of independent fields considered for each de Sitter solution in type IIB.}\label{tab:etaIIB+}
  \end{center}
\end{table}

A first observation is that $\eta_V < 0$ for all de Sitter solutions, in agreement with Conjecture 2 of \cite{Andriot:2019wrs}. This means that the solutions are unstable, and that a tachyon can be found among the fields $(\rho,\tau,\sigma_I)$ considered, in agreement with the proposal made in \cite{Danielsson:2012et}. While always successfully tested (see however \cite{Andriot:2021rdy} for a counter-example on a non-compact manifold), the check of this proposal is here extensive, since many different solution classes have been considered, including some (e.g.~$m_{46}^+$) where de Sitter solutions are found for the first time. We finally point out that for each solution, there is one and only one tachyonic mass in the spectrum.

A second observation is that most values are of order -1, in agreement with the refined de Sitter conjecture \cite{Garg:2018reu,Ooguri:2018wrx}. This is not surprising from the perspective of \cite{Andriot:2021rdy}, where it is argued that less generic stability behaviours need to be searched in specific corners of the parameter space, and here, we have not performed such dedicated searches. Our aim was rather to get (generic) solutions in many different classes.

Two exceptions are nevertheless worth being mentioned. The first one is $m_{46}^+ 5$, which admits a comparatively low value $|\eta_V| = 0.36462$. As indicated in Table \ref{tab:noncomp+}, the group manifold is however non-compact. The second one is $s_{6666}^+ 1$, which admits a comparatively high value $|\eta_V| = 18.445$. There, the group manifold is compact, see Table \ref{tab:comp+}. However, such a high instability is phenomenologically uninteresting.

Last but not least, let us add a word on the solution $s_{55}^+ 19$ found in \cite{Andriot:2021rdy}. Back then, its algebra was not identified. Thanks to the work of Section \ref{sec:algcomp}, we now know this algebra, and as indicated in Table \ref{tab:comp+}, the group manifold is compact. This is interesting, because this solution admits the lowest $|\eta_V|$ value known for a solution on a compact manifold: $\eta_V = - 0.12141$. This emphasizes the need for dedicated searches when it comes to stability of de Sitter solutions.

\subsubsection{Minkowski, and a new conjecture}\label{sec:stabMink}

For Minkowski solutions, we do not compute $\eta_V$ but look directly at the mass spectrum, provided in Appendix \ref{ap:spec}. Interestingly, we observe the \textsl{systematic presence of a massless mode}, in all solutions, the other masses being non-tachyonic. In solutions $m_{46}^0 2$ and $m_{466}^0 1-6$, there are even two massless modes. The systematic presence of such a 4d massless scalar field in classical, or at least 10d supergravity, Minkowski solutions is commonly believed to be true.\footnote{We thank T.~Van Riet for repeated support to this idea in private exchanges.} Examples are ubiquitous in the literature, a first one being Calabi-Yau compactifications (with $h^{1,1}\geq 1$). There, the presence of flat directions is related to the more general no-scale property of the potential \cite{Andrianopoli:2002vy, DallAgata:2013jtw}, which can in some models remove the dependence on some fields in the scalar potential. Less common examples include M-theory compactifications \cite{Blaback:2019zig}, compactifications to 6d \cite{Dibitetto:2019odu}, or maximal supergravity in 4d \cite{DallAgata:2012tne}, all having Minkowski solutions with massless scalar fields, some being flat directions. The systematic presence of a massless scalar field was even proven in compactifications to 4d ${\cal N}=1$ supergravity, coming from supersymmetric Minkowski solutions of 10d type IIA supergravity with certain $O_6/D_6$ \cite{Micu:2007rd, Ihl:2007ah} (see also \cite{deCarlos:2009qm}). This idea goes along with that of a systematic tachyon in de Sitter solutions.\footnote{Relations between the tachyon in a de Sitter solution and the sgoldstino in a (no-scale) Minkowski solution, the latter being the limit of the former, have been discussed in \cite{Covi:2008ea, Danielsson:2012et, Kallosh:2014oja, Junghans:2016uvg, Junghans:2016abx}. At first sight, we do not know whether our conjecture matches such a sgoldstino interpretation, but it would be interesting to investigate this further.} Following this line of thoughts and our observation, we propose here the following conjecture:
\bea
& \text{{\bf Massless Minkowski Conjecture:}} \label{conjMink}\\
& \nn\\[-10pt]
& \text{\textsl{10d supergravity solutions compactified to 4d Minkowski always admit a 4d massless scalar,}}\nn\\
& \text{\textsl{among the fields $(\rho,\tau,\sigma_I)$.}}\nn
\eea
\noindent The fact the massless mode should be among $(\rho,\tau,\sigma_I)$, and the claim not depending on supersymmetry (of the solution or of the 4d theory), are important additions with respect to previous related statements. The conjectured massless scalar field is also not necessarily a flat direction. These points make the conjecture more interesting, connecting directly to the proposal of \cite{Danielsson:2012et} stating a systematic de Sitter tachyon among the same fields. In addition, the complete set $(\rho,\tau,\sigma_I)$ is necessary: the massless mode is indeed not among $(\rho,\tau)$ alone in $s_{55}^0$ 1, $m_{46}^0$ 1,2, $m_{466}^0$ 1-6, as can be tested with {\tt MSSSp}; it is however in $s_{555}^0$ 1-4, probably because of the more limited supergravity contributions. Note that in heterotic string at order $\alpha'^0$, the field $\tau$ is massless in a Minkowski solution so the conjecture is valid, while fields $\sigma_I$ cannot be defined. Let us finally mention the recent apparent counter-example \cite{Bardzell:2022jfh}, where Minkowski solutions are found with all moduli stabilized. Those are however obtained on mirrors of rigid Calabi-Yau manifolds, which are better described as Landau-Ginzburg models, having $h^{1,1}=0$. As indicated there, since these models have no K\"ahler moduli, they do not have a proper 10d target space geometric description, and circumvent our conjecture by being not describable in 10d supergravity.\footnote{Similarly, without K\"ahler moduli, one cannot define internal volumes related to our $\rho$ and $\sigma_I$, and maybe not even the 4d dilaton $\tau$ which needs $\rho$. From this perspective, that example may even be viewed as being in agreement with the conjecture.}

An option would be to restrict the conjecture to solutions with 4d effective theories preserving at most ${\cal N}=1$ supersymmetry. Such a weaker statement could then be related to the Conjecture 4 of \cite{Andriot:2022way}, requiring at most ${\cal N}=1$ in the 4d effective theory for de Sitter solutions: the massless mode of Minkowski may then, once again, be related to the tachyon of de Sitter, both observed to be among $(\rho,\tau,\sigma_I)$. Nevertheless, preserving more supersymmetry typically corresponds to having less supergravity ingredients, leading to a simpler scalar potential, that would be less likely to generate a mass. So we stick to the above version of the conjecture. In addition, there exist examples of Minkowski solutions leading to a 4d theory with ${\cal N}\geq 2$ and having a massless scalar, starting with solution $s_{55}^0 1$ of \cite{Andriot:2020wpp} considered here in Appendix \ref{ap:spec}.

The conjecture applies in particular to classical Minkowski string backgrounds (see however below about corrections), and can as such get a swampland interpretation. Of course, it agrees with the anti-de Sitter distance conjecture \cite{Lust:2019zwm}, which provides in the asymptotics of field space a Minkowski solution with a massless mode coming from a tower. The conjecture \eqref{conjMink} is however stronger as it is not strictly about the asymptotics, and the massless mode is rather to be found among the light modes of the 4d theory (see a related discussion in Section \ref{sec:remscalesep}). A swampland-type corollary statement would then be the following:
\bea
& \text{{\bf Massless Minkowski Conjecture (swampland corollary):}} \label{conjMinkswamp}\\
& \nn\\[-10pt]
& \text{\textsl{In a quantum gravity 4d effective theory with a scalar potential $V(\phi^i)$, if a critical point}}\nn\\
& \text{\textsl{($\del_{\phi^j} V =0$) can be found in a region of field space corresponding to a classical and per-}}\nn\\
& \text{\textsl{turbative quantum gravity regime, and if this critical point is Minkowski ($V=0$), then}}\nn\\
& \text{\textsl{the mass matrix admits a vanishing eigenvalue.}}\nn
\eea
In addition, the conjecture \eqref{conjMink} specifies among which fields the massless mode can be found. Note that a vanishing mass matrix eigenvalue is equivalent to a degenerate Hessian of $V$. The above leads us to propose the following strong version of the conjecture:
\bea
& \text{{\bf Strong version:}} \label{conjMinkswampstrong}\\
& \nn\\[-10pt]
& \text{\textsl{If the above Minkowski critical point is realized, then there is no 4d tachyon, meaning}}\nn\\
& \hspace{1.6in} 0 = {\rm min}\ \nabla \del V = \frac{ V}{{M_p}^2} = \frac{|\nabla V|}{M_p} \ .\nn\\
& \text{\textsl{In other words, the inequalities of the refined de Sitter conjectures of \cite{Garg:2018reu,Ooguri:2018wrx,Andriot:2018mav} are}}\nn\\
& \text{\textsl{saturated.}}\nn
\eea
The strong version adds the information that the massless mode is the minimal eigenvalue of the mass matrix, meaning that there is no tachyon. This is indeed what we observe in our solutions.

There are two reasons to be careful about these swampland versions \eqref{conjMinkswamp} and \eqref{conjMinkswampstrong}. First, a quantum gravity effective theory would a priori contain many corrections going beyond the classical and perturbative regime. Even though they would be small in such a regime, there is no reason here (e.g.~without supersymmetry) for them to vanish. Any such non-vanishing correction could alter the claim of a vanishing mass. One should then be careful with the interpretation of the ``classical and perturbative regime'': whether this means a truncation of corrections (10d supergravity interpretation) or whether these are small, could change the statement. Second, we know that any additional scalar field with respect to our set $(\rho,\tau,\sigma_I)$ can a priori lower the value of ${\rm min}\ \nabla \del V$ (see below \eqref{etaV}). From this perspective, there is no reason for having no tachyon. In the literature, tachyons are however not observed in Minkowski compactifications (we do not consider here open string moduli, and e.g.~$D_p$-brane instabilities). So the strong version remains plausible. This conjecture deserves in any case more investigation, and we hope to come back to it in future work.\\

Contrary to other swampland conjectures related to stability, the conjecture \eqref{conjMink} does not depend on whether the solution is supersymmetric or not. Let us add here a word on this last question. The solutions found in \cite{Andriot:2022way} were obtained by solving the equations of motion and Bianchi identities. Conditions for supersymmetry, as e.g.~phrased in the language of generalized complex geometry with SU(3)$\times$SU(3) structures \cite{Grana:2006kf, Koerber:2010bx}, were not considered. Therefore, we see no reason for our solutions to be supersymmetric. For Minkowski solutions, a quick test goes as follows. Supersymmetric Minkowski solutions with $O_3$ typically need to have their $H$- and $F_3$-flux related through the ISD condition: $*_6 H= \epsilon\, g_s F_3$ \cite{Giddings:2001yu}, where for simplicity we do not specify the sign $\epsilon$ and we fix $e^{\phi}=g_s$. The class of Minkowski solutions with $O_p/D_p$ found in \cite{Andriot:2016ufg} generalises this relation to $*_{\bot} H^{(0)} = \epsilon\, g_s F^{(0)}_{6-p}$. The latter can be read in the smeared limit from the supersymmetry conditions as a particular solution, using the calibration condition $\iota^*[ 8\, {\rm Im} \Phi_2 ] = {\rm vol}_{||}$. Then, a hint for supersymmetry in a Minkowski solution is that appropriate components of $H$ and $g_s F_{6-p}$ take the same value. It is not the case in any of our solutions, except when both vanish. We conclude again that our solutions are unlikely to be supersymmetric.

\subsubsection{Anti-de Sitter}\label{sec:stabAdS}

The values of $\eta_V$ for each anti-de Sitter solution of \cite{Andriot:2022way} are given in Table \ref{tab:eta-}. We note already that all values satisfy $\eta_V \gtrsim -1$, in agreement with the conjecture of \cite{Gautason:2018gln} (see also Footnote \ref{foot:conjads}).

\begin{table}[H]
  \begin{center}
    \begin{tabular}{|c||c|c|c|c|c|c|c|c|c|}
    \hline
class & \multicolumn{4}{c||}{$s_{55}^-$} & \multicolumn{5}{c|}{$m_{46}^-$} \\
\hhline{-||----||-----}
solution & 1 & 2 & 3 & \multicolumn{1}{c||}{4} & 1 & 2 & 3 & 4 & 5 \\
    \hhline{=::====::=====}
$\eta_V$ & 0.7785 & -4 & -3.8495 & \multicolumn{1}{c||}{-2.4901} & 1.2531 & 1.5483 & 1.5537 & 1.3004 & 1.2548 \\
    \hline
    \end{tabular}
     \caption{Values of $\eta_V$ obtained with the set of fields considered for each anti-de Sitter solution.}\label{tab:eta-}
  \end{center}
\end{table}

The stability of anti-de Sitter solutions is more delicate. Let us first recall useful formulas valid for a 4d anti-de Sitter spacetime, extremum of a potential
\beq
\frac{{\cal R}_4}{4} = - \frac{3}{l^2} = \Lambda = \frac{V}{M_p^2} \ ,\label{relAdS}
\eeq
where $l$ is the so-called anti-de Sitter radius, appearing in the standard metric as follows $\d s^2 = \frac{l^2}{z^2} (\d z^2 + \d x_{\mu} \d x^{\mu} )$. Perturbative stability then requires for any scalar of mass $m$ to verify the Breitenlohner-Freedman (BF) bound, expressed in 4d as
\beq
m^2 > - \frac{9}{4\, l^2} \qquad \Rightarrow \quad \eta_V < \frac{3}{4} \ , \label{BF}
\eeq
from which we deduced an upper bound on $\eta_V$ in an anti-de Sitter solution. From this criterion, we see that all solutions with positive $\eta_V$ in Table \ref{tab:eta-} are perturbatively unstable.

Of interest are then the three anti-de Sitter solutions found with a negative $\eta_V$ (on compact group manifolds): not only those are perturbatively stable (at least within these fields), but their mass spectrum only has positive masses${}^2$. This perturbative stability may challenge to some extent the swampland conjecture on non-supersymmetric anti-de Sitter solutions \cite{Ooguri:2016pdq}, in case these solutions are non-supersymmetric. The latter is not straightforward to determine, and the quick test proposed for Minkowski solutions at the end of Section \ref{sec:stabMink} would not work for anti-de Sitter solutions, because of an extra term in the supersymmetry conditions, depending on the cosmological constant \cite{Koerber:2010bx}. However, as argued for Minkowski solutions, we still believe that our anti-de Sitter solutions are unlikely to be supersymmetric, making the above perturbative stability interesting.

Finally, we notice also the surprising values taken by $\eta_V$ in these perturbatively stable solutions. Of particular interest is $s_{55}^- 2$ which gets $\eta_V = -4.0000$ and $s_{55}^- 4$ with $\eta_V \approx -2.5$. The reason for such specific values might come from the particular field content of these solutions. Such choices for a solution ansatz may be of interest, and deserve more investigation. We will come back to these peculiar values in Section \ref{sec:remscalesep}.

\section{Scale separation}\label{sec:scalesep}

In this section we discuss the possibility of having scale separation in new anti-de Sitter solutions, found in previously unexplored solution classes $s_{55}^-$ and $m_{46}^-$ \cite{Andriot:2022way}. We also comment on a corresponding mass gap in Minkowski solutions. We first provide a general discussion and few observations in Section \ref{sec:remscalesep}. We then prove in Section \ref{sec:nogo} no-go theorems for anti-de Sitter solutions in $s_{55}^-$ and $m_{46}^-$ on nilmanifolds (including the torus) or manifolds with a Ricci flat metric, both argued in the Introduction to be relevant for scale separation.

\subsection{General comments on mass gap and scale separation}\label{sec:remscalesep}

As recalled in the Introduction, scale separation is a gap between the first non-zero mass of a tower of states (here taken as the Kaluza--Klein tower) and a 4d effective theory typical energy scale; such a gap then allows for an appropriate cut-off scale that truncates the tower. For anti-de Sitter, the 4d scale considered is given by the cosmological constant, while for Minkowski, it is set by the mass of light modes. To determine whether there is a scale separation with the first massive Kaluza--Klein state, one should access the latter scale. Beyond the torus, e.g.~on group manifolds, this is not an easy task: it typically requires to determine the eigenvalues of the Laplacian operator, as done e.g.~in \cite{Andriot:2016rdd, Andriot:2018tmb} for nilmanifolds. In particular, the first non-zero eigenvalue, of interest here, is not necessarily related to ${\cal R}_6$, the internal scalar curvature which sets another scale.

Beyond the Laplacian eigenvalues, another contribution to the mass of 4d modes is (the second derivative of) the scalar potential. In this paper, we only access the latter, and deduce from this potential our mass spectrum, displayed in Appendix \ref{ap:spec}. In addition, we only consider scalar fields with a dependence on 4d coordinates, i.e.~our truncation could be viewed as limited to the zero-modes of Kaluza--Klein towers. Their vanishing masses then get corrected by the scalar potential contribution: such fields are typically thought of as light modes. This interpretation is at least valid on a Ricci flat 6d manifold; a more careful analysis might be necessary here on group manifolds. Still, from this point of view, the mass spectrum we have at hand should not allow us to identify any scale separation. In our perturbatively stable anti-de Sitter solutions, $s_{55}^-$2-4, this seems consistent with the fact we do not observe important hierarchies between the masses${}^2$ and ${\cal R}_4$. In particular, $|\eta_V|$ is of order 1 (see Table \ref{tab:eta-}).\footnote{\label{foot:conjads}This agrees with the anti-de Sitter conjecture of \cite{Gautason:2018gln} which compares the mass of light modes to the cosmological constant, analogously to the criterion on $\eta_V$ of the refined de Sitter conjecture \cite{Ooguri:2018wrx}. The former differs from considering the mass scale of a tower, and the discussion on scale separation of \cite{Lust:2019zwm}.}

Despite the fact that we may not access the right scales to discuss scale separation, we will provide in the following two hints, that would conclude on the absence of scale separation in the new anti-de Sitter solutions found in the classes $s_{55}^-$ and $m_{46}^-$. A first hint is about integer values of conformal dimensions that we discuss below, a second one is given by no-go theorems for anti-de Sitter solutions discussed in Section \ref{sec:nogo}. Prior to this, we will also say a word on Minkowski solutions.\\

As mentioned in the Introduction, so-called DGKT anti-de Sitter solutions, that we interpret as being part of $s_{6666}^-$, exhibit scale separation. Through the standard holographic correspondence, the light mode spectrum of these solutions with masses $m^2$ corresponds to dual CFT operators with conformal dimensions $\Delta$, via the relation
\beq
\Delta (\Delta - 3) = m^2 l^2 \ \ \Leftrightarrow\ \ \Delta_{\pm} = \frac{3}{2} \pm \frac{1}{2} \sqrt{9 + 4 m^2 l^2} \ ,
\eeq
where $l$ is the anti-de Sitter radius defined in \eqref{relAdS}. As first discussed in \cite{Conlon:2021cjk, Apers:2022zjx} and computed more generally in \cite{Apers:2022tfm}, supersymmetric DGKT solutions satisfy the surprising property that the $\Delta$ take integer values. As pointed out in \cite{Quirant:2022fpn}, it is also the case of some non-supersymmetric solutions, but not of all of them.

One may wonder whether this specificity of integer conformal dimensions is related to having scale separation, at least for some solutions of this class. If this would hold, one could simply test the light mode spectrum of other solutions: getting integers would at least be a hint of scale separation. For $\Delta$ being an integer, one gets, using \eqref{relAdS} at an anti-de Sitter extremum, the following first possible values
\beq
- M_p^2\, \frac{m^2}{V} \, =\, -\frac{2}{3} \,,\ 0 \,,\ \frac{4}{3} \,,\ \frac{10}{3} \,,\ 6 \,,\ \frac{28}{3} \,,\ \frac{40}{3} \,,\ 18 \,, \frac{70}{3} \,,\ \frac{88}{3} \,,\ ... \label{integerlist}
\eeq
We can then compare these numbers to our anti-de Sitter solutions found in new classes: none of them has a spectrum giving values close to the above. One could argue that we are considering a limited set of scalar fields, and adding more fields could alter our values, but we believe the modification would not be important. Let us also emphasize that some of the solutions were noticed in Section \ref{sec:stabAdS} to have integer or half integer values of $\eta_V$. These seemingly special values however do not match any entry of the list \eqref{integerlist}. Following this line of thoughts, one may conclude on the absence of scale separation in these new anti-de Sitter solutions.\\

Before presenting another argument, let us say a word on the new Minkowski solutions found in \cite{Andriot:2022way}. We already mentioned in Section \ref{sec:stabMink} the apparent systematic presence of a massless mode, from which we draw the Massless Minkowski Conjecture \eqref{conjMink}. We note in addition for some solutions, namely $s_{555}^0 1$, $m_{46}^0 1$ and $m_{466}^0 2,3,5$, the presence of a gap in the mass spectrum: see Appendix \ref{ap:spec}. The most important is in $m_{46}^0 1$: a ratio between two consecutive masses${}^2$ is 7390.9. While such a gap is important, it remains hard, as discussed above, to conclude anything with respect to the first massive mode of a tower. But these examples deserve more investigation, such as the study of the Laplacian spectrum. We note however that according to Table \ref{tab:0alg} and the discussion below, none of these gapped solutions were shown to be on a compact manifold, while compactness remains crucial in this discussion, e.g.~with respect to the Kaluza--Klein towers.

\subsection{No-go theorems for anti-de Sitter on Ricci flat or nilmanifolds}\label{sec:nogo}

As motivated in the Introduction, scale separation in anti-de Sitter solutions on group manifolds could be limited to those on nilmanifolds, including the torus, or manifolds with a Ricci flat metric. It is the case for the solutions found in the solution classes $s_{6666}^-$ \cite{DeWolfe:2005uu, Camara:2005dc, Acharya:2006ne, Marchesano:2019hfb, Cribiori:2021djm} and $m_{5577}^-$ \cite{Caviezel:2008ik, Petrini:2013ika, Cribiori:2021djm}. In \cite{Andriot:2022way}, two new solution classes with anti-de Sitter solutions on group manifolds were discovered, $s^-_{55}$ and $m^-_{46}$, sharing the same T-duality relation as the former two classes. We prove however in this section that anti-de Sitter solutions cannot be found in these classes on nilmanifolds, or manifolds with a Ricci flat metric, giving a hint against scale separation in these classes. We also compare this situation to that of the first two classes.\\

We start with the solution class $s_{55}$ with $O_5$ along 12, 34 and $D_5$ along 56. We first consider the 6d (trace-reversed) Einstein equation combined with the 4d Einstein equation \cite[(B.23) \& (B.24)]{Andriot:2022way}. We take the trace of the former along 56. Using the field content of that solution class \cite[(2.14)]{Andriot:2022way}, we obtain
\beq
2 \sum_{a,b=5,6} \delta^{ab} {\cal R}_{ab} = {\cal R}_4 + |H|^2 + g_s^2 \left( |F_1|^2 + |F_3|^2 + |F_5|^2 \right) + \frac{g_s}{3} \left( T_{10}^3 - T_{10} \right) \ ,
\eeq
where $T_{10}^3 \equiv T_{10}^{(5)_3} \leq 0 $ because it corresponds to the contributions of $D_5$ along 56. Using further Einstein traces and the dilaton e.o.m., namely \cite[(B.1) \& (B.22)]{Andriot:2022way}, to eliminate some fluxes, we get
\beq
2 \sum_{a,b=5,6} \delta^{ab} {\cal R}_{ab} -  2{\cal R}_6 = 2{\cal R}_4  + \frac{g_s}{3} T_{10}^3  \ . \label{eqs55}
\eeq
One has
\beq
2 \sum_{a,b=5,6} \delta^{ab} {\cal R}_{ab} - 2 {\cal R}_6 = - 2 \sum_{a,b=1}^{4} \delta^{ab} {\cal R}_{ab} \ .
\eeq
The field content of $s_{55}$ indicates that all structure constants have one index which is 5 or 6. Therefore, using the Ricci tensor on a group manifold
\beq
2\ {\cal R}_{cd} = - f^b{}_{ac} f^a{}_{bd} - \delta^{bg} \delta_{ah} f^h{}_{gc} f^a{}_{bd} + \frac{1}{2} \delta^{ah}\delta^{bj}\delta_{ci}\delta_{dg} f^i{}_{aj} f^g{}_{hb} \ , \label{Ricci}
\eeq
we obtain
\bea
2 \sum_{a,b=5,6} \delta^{ab} {\cal R}_{ab} - 2 {\cal R}_6 & = \sum_{a,b=1}^{6} \sum_{c,d=1}^{4} \left( \delta^{cd} f^b{}_{ac} f^a{}_{bd} + (f^a{}_{bd})^2 - \frac{1}{2} (f^d{}_{ab})^2 \right) \nn\\
& = \sum_{a,b=1}^{6} \sum_{c,d=1}^{4} \delta^{cd} f^b{}_{ac} f^a{}_{bd} +  \sum_{a=5,6} \sum_{b,d=1}^{4}   (f^a{}_{bd})^2 \label{fabcnilnogo} \ .
\eea
The first term in \eqref{fabcnilnogo} is a partial trace of the Killing form. The Killing form identically vanishes for nilmanifolds. In addition, a manifold with Ricci flat metric, i.e.~${\cal R}_{ab}=0$, has \eqref{fabcnilnogo} vanishing. We deduce
\beq
\mbox{Nilmanifold or Ricci flat in $s_{55}$}:\quad 2 \sum_{a,b=5,6} \delta^{ab} {\cal R}_{ab} -  2{\cal R}_6 \geq 0 \ .
\eeq
For an anti-de Sitter solution in $s_{55}^-$, the right-hand side of \eqref{eqs55} is however negative. This leads to a \textsl{no-go theorem on anti-de Sitter solutions in the class $s_{55}$ on nilmanifolds (including the torus) or manifolds with Ricci flat metric}. Interestingly, as can be seen in Table \ref{tab:comp-}, solutions $s_{55}^-$2-4 of \cite{Andriot:2022way} were found on the algebra $\mathfrak{g}_{3.5}^0 \oplus \mathfrak{g}_{3.5}^0$, which can lead to a solvmanifold with a Ricci flat metric (see e.g.~\cite{Andriot:2015sia}). Of course, it is not the case for these solutions, which have ${\cal R}_6 <0$.

We turn to the solution class $m_{46}$. It has $O_4$ along 4, $O_6$ along 123, and possible $D_6$ along 156, 256, 356. The contributions of the latter are denoted $T_{10}^{(6)_2}, T_{10}^{(6)_3}, T_{10}^{(6)_4}$ and are negative. We proceed as above, taking the trace along 56, to first get
\beq
2 \sum_{a,b=5,6} \delta^{ab} {\cal R}_{ab} = {\cal R}_4 + |H|^2 + g_s^2 \left( |F_2|^2 + |F_4|^2 \right) +  2 g_s \left( \frac{1}{7}( T_{10}^{(6)_2} + T_{10}^{(6)_3} + T_{10}^{(6)_4} ) - \sum_p \frac{T_{10}^{(p)}}{p+1} \right) \ ,\nn
\eeq
using that $F_0=F_6=0$ in this solution class, and then
\beq
2 \sum_{a,b=5,6} \delta^{ab} {\cal R}_{ab} -  2{\cal R}_6 = 2{\cal R}_4  + \frac{2}{7} g_s \left( T_{10}^{(6)_2} + T_{10}^{(6)_3} + T_{10}^{(6)_4}\right) \ . \label{eqm46}
\eeq
The field content of $m_{46}$ indicates that structure constants always have one index which is 5 or 6. We conclude as above
\beq
\mbox{Nilmanifold or Ricci flat in $m_{46}$}:\quad 2 \sum_{a,b=5,6} \delta^{ab} {\cal R}_{ab} -  2{\cal R}_6 \geq 0 \ ,
\eeq
and deduce from \eqref{eqm46} a \textsl{no-go theorem on anti-de Sitter solutions in the class $m_{46}$ on nilmanifolds (including a torus), or manifolds with a Ricci flat metric}.\footnote{As a side remark, one deduces the following constraints for Minkowski solutions
\bea
& \mbox{Minkowski solutions in $s_{55}$ on a nilmanifold}:\quad f^{5,6}{}_{bd} = 0 \ ,\ T_{10}^3=0 \ ,\\
& \hspace{-0.2in} \mbox{Mink. sol. in $m_{46}$ on a nilmanifold}:\quad f^{5,6}{}_{bd} = 0 \ ,\ T_{10}^{(6)_2} = T_{10}^{(6)_3} = T_{10}^{(6)_4}=0 \ ,\nn
\eea
leading to the conclusion that only two sets of sources can be present in either of those classes. This is consistent with our solutions in $s^0_{55}$ and $m^0_{46}$, and those already known.}

These no-go theorems are certainly consistent with our searches for solutions in $s^-_{55}$ and $m^-_{46}$. Whether or not they prevent from getting scale separation is not established, but as argued in the Introduction, this is possibly a relevant criterion. Let us finally compare to the situation in the other classes. Proceeding similarly for $m_{5577}$ with $O_5$ along 12, 34 and $O_7$ along 2456, 1356, we obtain the following equality
\beq
2 \sum_{a,b=5,6} \delta^{ab} {\cal R}_{ab} -  2{\cal R}_6 = 2{\cal R}_4  + \frac{g_s}{4} T_{10}^{(7)}  \ . \label{eqm5577}
\eeq
For the same reason as above, the left-hand side has to be positive or zero on nilmanifolds (including a torus) and manifolds with a Ricci flat metric. To avoid a no-go theorem for an anti-de Sitter solution in $m_{5577}^-$ on such a manifold, we deduce the requirement $ T_{10}^{(7)} = T_{10}^{(7)_1} + T_{10}^{(7)_2} > 0$. This means that the positive contribution of $O_7$ in those should not be dominated by that of possible $D_7$, negative. Getting such a requirement is interesting, but we also identify an important difference with $s^-_{55}$ and $m^-_{46}$:  the absence of directions with only $D_p$-branes. This difference is even stronger with $s_{6666}$ where we have difficulties identifying relevant directions over which to trace as above: doing so brings further contributions to the equations, leading to looser requirements, not worth being indicated here. \textsl{The presence or absence of directions with only $D_p$-branes is related this way to the possibility of getting anti-de Sitter solutions on Ricci flat or nilmanifolds, which in turn could be related to scale separation.} These relations deserve more investigation.

\vfill

\subsection*{Acknowledgements}

We thank H.~Skarke and D.~Tsimpis for helpful exchanges during the completion of this work. P.~M.~thanks the ITP at TU Wien for hospitality and for the opportunity to work on this project. L.~H.~acknowledges support from the Austrian Science Fund (FWF): project number P34562-N, doctoral program W1252-N27.

\newpage

\begin{appendix}

\section{Subtleties on the parameters in solvable algebras}\label{ap:par}

Real 6d indecomposable unimodular solvable Lie algebras are classified in \cite{Bock} according to their nilradical, into so-called ``isomorphism classes''. This means that for any such algebra, an isomorphism can be found that maps it to one (and only one) of these classes. Some of these classes, as presented in \cite{Bock}, however depend on continuous parameters: for instance, $\mathfrak{g}_{6.93}^{0,\nu_0}$ depends on $\nu_0$. These real parameters sometimes take values in a certain range, to which we come back below. Contrary to what one would expect from the name ``isomorphism class'', one can actually not set the parameter to any fixed value with an isomorphism; in other words, one ``class'' with a parameter actually corresponds to an infinite number of non-isomorphic algebras. Another interpretation is to view $\mathfrak{g}_{6.93}^{0,\nu_0}$ as a different ``isomorphism class'' for every (allowed) value of the continuous parameter $\nu_0$. This subtlety helps understanding that properties such as the Killing form signature or the CS and DS, meant to be basis independent, can actually depend (and change) with the continuous parameter: see Table \ref{tab:algsolv}. The nilradical however does not change, in agreement with the classification.

Everything just written holds given a properly specified allowed range of the parameters. Such a range is however rarely given in \cite{Bock}, and this leads to a few issues that we now mention. In some cases, special values of the parameters actually allow to have them set to a fixed value thanks to an isomorphism. This hints at a possible better characterisation of their range preventing any such issue. For example, we consider $\mathfrak{g}_{6.88}^{0,\mu_0,\nu_0}$ in the case where $\nu_0=0$, $\mu_0 \neq 0$. In that case, a rescaling allows to bring $\mu_0$ to 1, meaning that it is not a true parameter anymore. In other words, for $\mu_0 \neq 0$, $\mathfrak{g}_{6.88}^{0,\mu_0,0}= \mathfrak{g}_{6.88}^{0,1,0}$. This seems to be generalizable beyond the case $\nu_0=0$.

Another issue is the following. When computing the CS for the algebras of Table \ref{tab:algsolv} but following \cite{Bock}, we noticed that for some parameter values, the CS would reach 0. This implies that the algebra is nilpotent, which cannot be the case since the algebra is 6-dimensional and the nilradical is 5-dimensional. Based on this, we indicated ourselves restrictions (i.e.~the range) on the parameters in Table \ref{tab:algsolv}. In Table \ref{tab:algsolvnil}, we indicate what the CS and DS would be if we allow for different values of the parameters, and the corresponding nilpotent algebra.

\begin{table}[H]
  \begin{center}
    \begin{tabular}{|c|c|c|c|c|}
    \hline
Algebra & Parameters & CS & DS & Nilpotent algebra \\
\hhline{=====}
$\mathfrak{g}_{6.83}^{0,l}$ & $l=0$ & 6,3,1,0 & 6,3,0 & $\mathfrak{g}_{6.N14}^{-1}$ \\
\hline
$\mathfrak{g}_{6.88}^{0,\mu_0,\nu_0}$ & $\mu_0 = \nu_0 = 0$ & 6,1,0 & 6,1,0 & $\mathfrak{g}_{5.4} \oplus \mathfrak{u}(1)$ \\
\hline
$\mathfrak{g}_{6.89}^{0,\nu_0,s}$ & $s = \nu_0 = 0$ & 6,1,0 & 6,1,0 & $\mathfrak{g}_{5.4} \oplus \mathfrak{u}(1)$ \\
\hline
$\mathfrak{g}_{6.92}^{0,\mu_0,\nu_0}$ & $\mu_0\nu_0 = 0, |\mu_0| + |\nu_0| \neq 0$ & 6,3,1,0 & 6,3,0 & $\mathfrak{g}_{6.N14}^{-1}$ \\
 & $\mu_0 = \nu_0 = 0$ & 6,1,0 & 6,1,0 & $\mathfrak{g}_{5.4} \oplus \mathfrak{u}(1)$ \\
\hline
    \end{tabular}
     \caption{For some algebras of Table \ref{tab:algsolv}, further values for the parameters and corresponding CS and DS. The CS reaching 0, one concludes on nilpotent algebras, so the parameter values cannot be allowed. We give the corresponding nilpotent algebra, with notations of \cite{Bock}.}\label{tab:algsolvnil}
  \end{center}
\end{table}

\section{An analytical change of basis}\label{ap:chgebasis}

We present here a general analytical change of basis for solutions $m^+_{46}4,5$, identifying the algebra to be non-solvable. The Table \ref{tab:alg} could have been enough for this identification, but this explicit change of basis could serve further purposes.

The solutions $m^+_{46}4,5$ have the following non-zero structure constants
\bea
\label{scm46:4}
m^+_{46}4,5: \quad & {f^1}_{45},{f^1}_{46},{f^2}_{45},{f^2}_{46},{f^3}_{45},{f^3}_{46},{f^4}_{15},{f^4}_{16}, {f^4}_{25},{f^4}_{26},{f^4}_{35},{f^4}_{36},\nn \\&
{f^5}_{14},{f^5}_{24},{f^5}_{34},{f^6}_{14},{f^6}_{24},{f^6}_{34}\, .
\eea
We first perform the following change of basis
\beq
e^{a\neq 6'}=e^a, e^{6'}=e^6+\frac{{f^4}_{25}}{{f^4}_{26}} e^5\, .
\eeq
Using the Jacobi identities $ f^{a}{}_{e[6}f^e{}_{b5]} = 0 $, $a,b=1,2,3$, one can show that the above ratio enters the following equalities
\beq
\frac{{f^4}_{b5}}{{f^4}_{b6}} = \frac{{f^{a}}_{45}}{{f^{a}}_{46}} \ ,\ a,b=1,2,3 \ .
\eeq
This allows to reduce the set of structure constants to the following
\bea
\label{scm46:4p}
& {{f^1}_{46}}',{{f^2}_{46}}',{{f^3}_{46}}',{{f^4}_{16}}',{{f^4}_{26}}',{{f^4}_{36}}',{{f^5}_{14}}',{{f^5}_{24}}',{{f^5}_{34}}':\ {\rm unchanged}\nn \\&
{{f^6}_{a4}}'={f^6}_{a4}+ \frac{{f^4}_{25}}{{f^4}_{26}} {f^5}_{a4}\ ,\ a=1,2,3\ .
\eea
We now consider the Jacobi identities ${f^{5,6}{}_{e[a}f^e{}_{b6]}}'=0$, for $a,b$ taking values among $1,2,3$. From those we deduce the following equalities
\beq
\frac{{f^4{}_{a6}}'}{{f^4{}_{b6}}'}= \frac{{f^6{}_{a4}}'}{{f^6{}_{b4}}'} = \frac{{f^5{}_{a4}}'}{{f^5{}_{b4}}'}\ \Rightarrow\ \frac{{{f^6}_{14}}'}{{{f^5}_{14}}'}= \frac{{{f^6}_{24}}'}{{{f^5}_{24}}'}= \frac{{{f^6}_{34}}'}{{{f^5}_{34}}'}=\tilde{\alpha} \ .
\eeq
We then perform the change of basis
\beq
\label{m46:4cb1}
e^{a\neq 5''}=e^{a'},\ e^{5''}=e^{6'}- \tilde{\alpha} e^{5'}\, .
\eeq
This allows to set to zero the ${{f^5}_{a4}}''$ without changing the others. We are left with
\bea
\label{scm46:4new}
& {{f^1}_{46}}'',{{f^4}_{16}}'',{{f^2}_{46}}'',{{f^4}_{26}}'',{{f^3}_{46}}'',{{f^4}_{36}}'':\ {\rm unchanged}\nn \\&
{{f^6}_{a4}}''={f^6}_{a4}+ \frac{{f^4}_{25}}{{f^4}_{26}} {f^5}_{a4}\ ,\ a=1,2,3\ \mbox{(i.e.~unchanged)}\ .
\eea
To reduce further the number of structure constants, we consider the general transformation
\beq
\label{m46:1cb3}
e^{a\neq 1''',3'''}=e^{a''}, e^{1'''}= e^{1''}-\beta e^{3''}, e^{3'''}= e^{1''}+\delta e^{3''}\ ,\ \beta + \delta \neq 0 \ .
\eeq
With $\beta = \frac{{{f^1}_{46}}''}{{{f^3}_{46}}''}$, $\delta = \frac{{{f^4}_{36}}''}{{{f^4}_{16}}''}$, this allows us to get rid of one pair of structure constants, namely set to zero ${{f^1}_{46}}''' ={{f^4}_{16}}'''=0$, while some of the new structure constants are
\beq
\label{scm46newnew}
{{f^2}_{46}}'''= {{f^2}_{46}}'' \ ,\ {{f^4}_{26}}''' = {{f^4}_{26}}''\ ,\ {{f^3}_{46}}'''= {{f^1}_{46}}'' + {{f^3}_{46}}'' \frac{{{f^4}_{36}}''}{{{f^4}_{16}}''}\ ,\ {{f^4}_{36}}'''= {{f^4}_{16}}'' \ .
\eeq
Verifying $\beta+\delta \neq 0$ amounts here to
\beq
{{f^3}_{46}}'' {{f^4}_{16}}'' = {{f^3}_{46}} {{f^4}_{16}} \neq 0 \ ,\  {{f^1}_{46}}'' {{f^4}_{16}}'' + {{f^3}_{46}}'' {{f^4}_{36}}'' =  {{f^1}_{46}} {{f^4}_{16}} + {{f^3}_{46}} {{f^4}_{36}} \neq 0 \ , \label{betadelta1}
\eeq
which is satisfied in our solutions. We should still determine the resulting ${{f^6}_{a4}}'''$. The Jacobi identity ${f^6{}_{e[1}}''{f^e{}_{36]}}''=0$ gives us (as above) the equality ${f^6{}_{14}}''{f^4{}_{36}}''= {f^4{}_{16}}'' {f^6{}_{34}}''$. This allows us first to verify that ${{f^6}_{14}}'''=0$. We obtain in addition
\beq
{{f^6}_{34}}'''= {{f^4}_{16}}''  \frac{ {{f^1}_{46}}'' {{f^6}_{14}}'' + {{f^3}_{46}}'' {{f^6}_{34}}'' }{ {{f^1}_{46}}'' {{f^4}_{16}}'' + {{f^3}_{46}}'' {{f^4}_{36}}''} = {{f^6}_{14}}'' \ ,
\eeq
which gets simplified as indicated thanks again to the above Jacobi identity. We eventually get the following remaining structure constants
\bea
\label{scm46:4newp}
& {{f^2}_{46}}''',{{f^4}_{26}}''',{{f^6}_{24}}''':\ {\rm unchanged}\nn \\&
{{f^3}_{46}}'''={{f^1}_{46}}'' + {{f^3}_{46}}'' \frac{{{f^4}_{36}}''}{{{f^4}_{16}}''} \ ,\ {{f^4}_{36}}'''= {{f^4}_{16}}'' \ ,\ {{f^6}_{34}}'''= {{f^6}_{14}}''   \ .
\eea
We can then iterate the process, considering
\beq
\label{m46:1cb4}
e^{a\neq 2'''',3''''}=e^{a'''}, e^{2''''}= e^{2'''}-\beta e^{3'''}, e^{3''''}= e^{2'''}+\delta e^{3'''}\ ,\ \beta + \delta \neq 0 \ .
\eeq
Verifying $\beta+\delta \neq 0$ means
\beq
{{f^3}_{46}}''' {{f^4}_{26}}''' \neq 0 \ ,\  {{f^2}_{46}}''' {{f^4}_{26}}''' + {{f^3}_{46}}''' {{f^4}_{36}}'''  \neq 0 \ ,
\eeq
which can be reformulated, given \eqref{betadelta1} and that ${{f^4}_{26}} \neq 0$ was already required in the first transformation, as
\beq
 {{f^2}_{46}} {{f^4}_{26}} +  {{f^1}_{46}} {{f^4}_{16}} + {{f^3}_{46}} {{f^4}_{36}} \neq 0 \ .
\eeq
Once again, the Jacobi identity ${f^6{}_{e[2}}'''{f^e{}_{36]}}'''=0$ gives us the equality ${f^6{}_{24}}'''{f^4{}_{36}}'''= {f^4{}_{26}}''' {f^6{}_{34}}'''$ which can be used to simplify the results. We obtain the following remaining structure constants
\bea
\label{scm46:4newpp}
& {{f^3}_{46}}''''={{f^2}_{46}}''' + {{f^3}_{46}}''' \frac{{{f^4}_{36}}'''}{{{f^4}_{26}}'''} \ ,\ {{f^4}_{36}}''''= {{f^4}_{26}}''' \ ,\ {{f^6}_{34}}''''=  {{f^6}_{24}}''' \ ,
\eea
or simplified
\bea
\label{scm46:4final}
& {{f^3}_{46}}''''=  \frac{1}{{f^4}_{26}} \left(  {f^2}_{46} {f^4}_{26} +   {{f^1}_{46}} {{f^4}_{16}} + {{f^3}_{46}} {{f^4}_{36}}  \right) \ ,\ {{f^4}_{36}}''''= {f^4}_{26} \nn\\&
{{f^6}_{34}}''''= {f^6}_{24}+ \frac{{f^4}_{25}}{{f^4}_{26}} {f^5}_{24}  \ .
\eea
This eventually gives the numerical values
\bea
& m^+_{46}4: \quad {{f^3}_{46}}''''=-0.0871207\ ,\ {{f^4}_{36}}''''=-0.628272\ ,\ {{f^6}_{34}}''''=1.70966\, ,\nn\\
& m^+_{46}5: \quad {{f^3}_{46}}''''=-2.13542\ ,\ {{f^4}_{36}}''''=-0.0047329\ ,\ {{f^6}_{34}}''''=0.00142542\, .\label{scm46:4newnewnew}
\eea
The relative signs allow us to identify the algebras to be all the same
\beq
m^+_{46}4,5:\mathfrak{so}(2,1)\oplus3\, \mathfrak{u}(1)\, .\label{scm46:4algebra}
\eeq

Interestingly the same procedure to get the change of basis can be applied to solutions $m^+_{46}6$ and $m^+_{46}8$, which admit as starting structure constants a subset of the above ones \eqref{scm46:4}.

\section{Independent fields and field space metric}\label{ap:gij}

In Section \ref{sec:met}, we explained how to identify a set of independent fields and compute the associated non-degenerate field space metric. We give in the following this data for each case encountered in our solutions, beyond the example given in \eqref{gijm461}. We present the source sets ordered, the corresponding independent scalar fields, and the field space metric expressed in that field basis.

\hspace{0.1in}

$\boldsymbol{s_{55}}$: $O_5$ (12, 34), $D_5$ (56), or $\boldsymbol{s_{555}}$: $O_5$ (12,34,56). Fields: $(\rho, \tau, \sigma_1, \sigma_2)$
\begin{equation}
g_{ij} = M_p^2\,
\begin{pmatrix}
\mathlarger{\frac{3}{2 \rho^2}} & 0 & 0 &0\\[10pt]
0 & \mathlarger{\frac{2}{\tau^2}} & 0 &0 \\[10pt]
0 & 0 & \mathlarger{\frac{12}{\sigma_1^2}} &\mathlarger{-\frac{6}{\sigma_1 \sigma_2}} \\[10pt]
0 & 0 &\mathlarger{-\frac{6}{\sigma_1 \sigma_2}} & \mathlarger{\frac{12}{\sigma_2^2}}
\end{pmatrix} \ . \label{gijs55}
\end{equation}

$\boldsymbol{s_{66}}$: $O_6$ (123, 145), $D_6$ (256, 346), or $\boldsymbol{s_{6666}}$: $O_6$ (123, 145, 256, 346). Fields: $(\rho, \tau, \sigma_1, \sigma_2, \sigma_3)$
\beq
g_{ij} = M_p^2\,
\begin{pmatrix}
\mathlarger{\frac{3}{2 \rho^2}} & 0 & 0 &0 & 0  \\[10pt]
0 & \mathlarger{\frac{2}{\tau^2}} & 0 &0 &0 \\[10pt]
0 & 0 & \mathlarger{\frac{27}{2 \sigma_1^2}} & \mathlarger{-\frac{9}{2\sigma_1 \sigma_2}} & \mathlarger{-\frac{9}{2\sigma_1 \sigma_3}} \\[10pt]
0 & 0 &\mathlarger{-\frac{9}{2\sigma_1 \sigma_2}} & \mathlarger{\frac{27}{2\sigma_2^2}} & \mathlarger{-\frac{9}{2\sigma_2 \sigma_3}}  \\[10pt]
0 & 0 &\mathlarger{-\frac{9}{2\sigma_1 \sigma_3}} & \mathlarger{-\frac{9}{2\sigma_2 \sigma_3}} & \mathlarger{\frac{27}{2\sigma_3^2}}
\end{pmatrix} \ .\label{gijs6666}
\eeq

$\boldsymbol{m_{46}}$ (3 $D_6$): $O_4$ (4), $O_6$ (123), $D_6$ (156, 256, 356). Fields: $(\rho, \tau, \sigma_1, \sigma_2, \sigma_3, \sigma_4)$
\begin{equation}
g_{ij} = M_p^2\,
\begin{pmatrix}
\mathlarger{\frac{3}{2 \rho^2}} & 0 & 0 &0 & 0 & 0 \\[10pt]
0 & \mathlarger{\frac{2}{\tau^2}} & 0 &0 &0 &0 \\[10pt]
0 & 0 & \mathlarger{\frac{15}{2 \sigma_1^2}} & \mathlarger{-\frac{9}{2\sigma_1 \sigma_2}} & \mathlarger{-\frac{9}{2\sigma_1 \sigma_3}} & \mathlarger{-\frac{9}{2\sigma_1 \sigma_4}} \\[10pt]
0 & 0 &\mathlarger{-\frac{9}{2\sigma_1 \sigma_2}} & \mathlarger{\frac{27}{2\sigma_2^2}} & \mathlarger{-\frac{9}{2\sigma_2 \sigma_3}} & \mathlarger{-\frac{9}{2\sigma_2 \sigma_4}} \\[10pt]
0 & 0 &\mathlarger{-\frac{9}{2\sigma_1 \sigma_3}} & \mathlarger{-\frac{9}{2\sigma_2 \sigma_3}} & \mathlarger{\frac{27}{2\sigma_3^2}} & \mathlarger{\frac{9}{2\sigma_3 \sigma_4}} \\[10pt]
0 & 0 &\mathlarger{-\frac{9}{2\sigma_1 \sigma_4}} & \mathlarger{-\frac{9}{2\sigma_2 \sigma_4}} & \mathlarger{\frac{9}{2\sigma_3 \sigma_4}} & \mathlarger{\frac{27}{2\sigma_4^2}}
\end{pmatrix} \ . \label{gijm462}
\end{equation}

$\boldsymbol{m_{55}}$: $O_5$ (12, 34), $D_5$ (56), $D_7$ (2456, 2356, 1456, 1356), or $O_5$ (12, 34), $D_7$ (2456, 2356, 1456, 1356), or $O_5$ (12, 34), $D_7$ (2456, 1456, 1356). Fields: $(\rho, \tau, \sigma_1, \sigma_2, \sigma_4, \sigma_6)$
\begin{equation}
g_{ij} = M_p^2\,
\begin{pmatrix}
\mathlarger{\frac{3}{2 \rho^2}} & 0 & 0 &0 &0 &0\\[10pt]
0 & \mathlarger{\frac{2}{\tau^2}} & 0 &0 &0 &0\\[10pt]
0 & 0 & \mathlarger{\frac{12}{\sigma_1^2}} &\mathlarger{-\frac{6}{\sigma_1 \sigma_2}} & \mathlarger{-\frac{3}{\sigma_1 \sigma_4}} & \mathlarger{-\frac{3}{\sigma_1 \sigma_6}} \\[10pt]
0 & 0 &\mathlarger{-\frac{6}{\sigma_1 \sigma_2}} & \mathlarger{\frac{12}{\sigma_2^2}} & \mathlarger{-\frac{3}{\sigma_2 \sigma_4}} & \mathlarger{-\frac{3}{\sigma_2 \sigma_6}} \\[10pt]
0 & 0 &\mathlarger{-\frac{3}{\sigma_1 \sigma_4}} & \mathlarger{-\frac{3}{\sigma_2 \sigma_4}} & \mathlarger{\frac{12}{\sigma_4^2}} & \mathlarger{\frac{3}{\sigma_4 \sigma_6}} \\[10pt]
0 & 0 &\mathlarger{-\frac{3}{\sigma_1 \sigma_6}} & \mathlarger{-\frac{3}{\sigma_2 \sigma_6}} & \mathlarger{\frac{3}{\sigma_4 \sigma_6}} & \mathlarger{\frac{12}{\sigma_6^2}} \\[10pt]
\end{pmatrix} \ . \label{gijm55}
\end{equation}

$\boldsymbol{m_{5577}}$: $O_5$ (12, 34), $O_7$ (1356, 2456), or $\boldsymbol{m_{5577}^*}$: $O_5$ (12, 34), $O_7$ (1456, 2356).

Fields: $(\rho, \tau, \sigma_1, \sigma_2, \sigma_3)$
\begin{equation}
g_{ij} = M_p^2\,
\begin{pmatrix}
\mathlarger{\frac{3}{2 \rho^2}} & 0 & 0 &0 &0\\[10pt]
0 & \mathlarger{\frac{2}{\tau^2}} & 0 &0 &0\\[10pt]
0 & 0 & \mathlarger{\frac{12}{\sigma_1^2}} &\mathlarger{-\frac{6}{\sigma_1 \sigma_2}} & \mathlarger{-\frac{3}{\sigma_1 \sigma_3}} \\[10pt]
0 & 0 &\mathlarger{-\frac{6}{\sigma_1 \sigma_2}} & \mathlarger{\frac{12}{\sigma_2^2}} & \mathlarger{-\frac{3}{\sigma_2 \sigma_3}} \\[10pt]
0 & 0 &\mathlarger{-\frac{3}{\sigma_1 \sigma_3}} & \mathlarger{-\frac{3}{\sigma_2 \sigma_3}} & \mathlarger{\frac{12}{\sigma_3^2}} &
\end{pmatrix} \ . \label{gijm5577}
\end{equation}

\section{Mass spectrum of Minkowski and anti-de Sitter solutions}\label{ap:spec}

We provide in this appendix few information on the Minkowski and anti-de Sitter solutions found in \cite{Andriot:2022way}; more can be found in that reference. We also consider $s_{55}^0 1$ found in \cite{Andriot:2020wpp}. For each solution, we give the 4d Ricci scalar ${\cal R}_4 $, the 6d one ${\cal R}_6$ and the mass spectrum. For anti-de Sitter solutions, we give in addition the value of the parameter $\eta_V$. Definitions and comments on these quantities can be found in sections \ref{sec:stab} and \ref{sec:scalesep}. The numerical values are given in units of $2\pi l_s$.

\subsection{Minkowski solutions}

\subsection*{$\boldsymbol{s_{55}^0 1}$}

\begin{equation*}
{\cal R}_4 = 0  \,, \quad  {\cal R}_6 = -1.0206 \,,
\end{equation*}
\begin{equation*}
\text{masses}^2 = (3.6377, 1.5406, 0.33559, 0)  \,.
\end{equation*}

\subsection*{$\boldsymbol{s_{555}^0 1}$}

\begin{equation*}
{\cal R}_4 = 0  \,, \quad  {\cal R}_6 =-0.017241 \,,
\end{equation*}
\begin{equation*}
\text{masses}^2 = (0.052928,0.0021215,0.00005291,0) \,.
\end{equation*}

\subsection*{$\boldsymbol{s_{555}^0 2}$}

\begin{equation*}
{\cal R}_4 = 0  \,, \quad  {\cal R}_6 =-0.11649 \,,
\end{equation*}
\begin{equation*}
\text{masses}^2 = (0.83127,0.07301,0.068032,0) \,.
\end{equation*}

\subsection*{$\boldsymbol{s_{555}^0 3}$}

\begin{equation*}
	{\cal R}_4 = 0 \,, \quad  {\cal R}_6 =-0.14383 \,,
\end{equation*}
\begin{equation*}
\text{masses}^2 = (0.2163,0.098852,0.045967,0) \,.
\end{equation*}

\subsection*{$\boldsymbol{s_{555}^0 4}$}

\begin{equation*}
	{\cal R}_4 = 0  \,, \quad  {\cal R}_6 =-0.11298 \,,
\end{equation*}
\begin{equation*}
\text{masses}^2 = (0.27831,0.077819,0.032095,0) \,.
\end{equation*}

\subsection*{$\boldsymbol{m_{46}^0 1}$}

\begin{equation*}
{\cal R}_4 = 0 \,, \quad  {\cal R}_6 = -0.015368 \,,
\end{equation*}
\begin{equation*}
\text{masses}^2 = (3.3631, 0.45394, 0.067729, 9.1638 \cdot 10^{-6}, 0) \,.
\end{equation*}

\subsection*{$\boldsymbol{m_{46}^0 2}$}

\begin{equation*}
{\cal R}_4 = 0 \,, \quad  {\cal R}_6 = -0.023897 \,,
\end{equation*}
\begin{equation*}
\text{masses}^2 = (0.52608, 0.077079, 0.021226, 0, 0) \,.
\end{equation*}

\subsection*{$\boldsymbol{m_{466}^0 1}$}

\begin{equation*}
{\cal R}_4 = 0 \,, \quad  {\cal R}_6 = -0.026276 \,,
\end{equation*}
\begin{equation*}
\text{masses}^2 = (0.26972, 0.074729, 0.020261, 0, 0) \,.
\end{equation*}

\subsection*{$\boldsymbol{m_{466}^0 2}$}

\begin{equation*}
{\cal R}_4 = 0 \,, \quad  {\cal R}_6 = -0.039542 \,,
\end{equation*}
\begin{equation*}
\text{masses}^2 = (0.23513, 0.03448, 0.00023868, 0, 0) \,.
\end{equation*}

\subsection*{$\boldsymbol{m_{466}^0 3}$}

\begin{equation*}
{\cal R}_4 = 0 \,, \quad  {\cal R}_6 = -0.00043667 \,,
\end{equation*}
\begin{equation*}
\text{masses}^2 = (0.026127, 0.015642, 0.00062489, 0, 0) \,.
\end{equation*}

\subsection*{$\boldsymbol{m_{466}^0 4}$}

\begin{equation*}
{\cal R}_4 = 0 \,, \quad  {\cal R}_6 = -0.036741  \,,
\end{equation*}
\begin{equation*}
\text{masses}^2 = (0.17069, 0.012707, 0.0044701, 0, 0) \,.
\end{equation*}

\subsection*{$\boldsymbol{m_{466}^0 5}$}

\begin{equation*}
{\cal R}_4 = 0 \,, \quad  {\cal R}_6 = -0.034908 \,,
\end{equation*}
\begin{equation*}
\text{masses}^2 = (0.32049, 0.11059, 0.0073101, 0, 0) \,.
\end{equation*}

\subsection*{$\boldsymbol{m_{466}^0 6}$}

\begin{equation*}
{\cal R}_4 = 0 \,, \quad  {\cal R}_6 = -0.059001 \,,
\end{equation*}
\begin{equation*}
\text{masses}^2 = (0.21201, 0.035651, 0.013395, 0, 0) \,.
\end{equation*}

\subsection{Anti-de Sitter solutions}

\subsection*{$\boldsymbol{s_{55}^- 1}$}

\begin{equation*}
	{\cal R}_4 = -0.033561  \,, \quad  {\cal R}_6 =-0.0073162 \,, \quad \eta_V= 0.7785 \,,
\end{equation*}
\begin{equation*}
\text{masses}^2 = (0.19854,0.060726,0.04147,-0.0065318) \,.
\end{equation*}

\subsection*{$\boldsymbol{s_{55}^- 2}$}

\begin{equation*}
	{\cal R}_4 = -0.015208  \,, \quad  {\cal R}_6 =-0.017287 \,, \quad \eta_V= -4 \,,
\end{equation*}
\begin{equation*}
\text{masses}^2 = (0.070021,0.044657,0.027383,0.015208)  \,.
\end{equation*}

\subsection*{$\boldsymbol{s_{55}^- 3}$}

\begin{equation*}
	{\cal R}_4 = -0.036862  \,, \quad  {\cal R}_6 =-0.023797 \,, \quad \eta_V= -3.8495  \,,
\end{equation*}
\begin{equation*}
\text{masses}^2 = (0.20393,0.11596,0.074406,0.035475) \,.
\end{equation*}

\subsection*{$\boldsymbol{s_{55}^- 4}$}

\begin{equation*}
	{\cal R}_4 = -0.024424  \,, \quad  {\cal R}_6 =-0.023691 \,, \quad \eta_V=-2.4901  \,,
\end{equation*}
\begin{equation*}
\text{masses}^2 = (0.15904,0.067206,0.039032,0.015205 )\,.
\end{equation*}

\subsection*{$\boldsymbol{m_{46}^- 1}$}

\begin{equation*}
{\cal R}_4 = -0.048164 \,, \quad  {\cal R}_6 = -0.02412 \,, \quad \eta_V= 1.2531\,,
\end{equation*}
\begin{equation*}
\text{masses}^2 = (0.49918, 0.13392, 0.060085, 0.054407, -0.015089) \,.
\end{equation*}

\subsection*{$\boldsymbol{m_{46}^- 2}$}

\begin{equation*}
{\cal R}_4 = -0.019002 \,, \quad  {\cal R}_6 = -0.012892 \,, \quad \eta_V=  1.5483 \,,
\end{equation*}
\begin{equation*}
\text{masses}^2 = (0.38901, 0.18817, 0.031941, 0.013066, -0.0073556) \,.
\end{equation*}

\subsection*{$\boldsymbol{m_{46}^- 3}$}

\begin{equation*}
{\cal R}_4 = -0.1534 \,, \quad  {\cal R}_6 = -0.11122 \,, \quad \eta_V= 1.5537 \,,
\end{equation*}
\begin{equation*}
\text{masses}^2 = (3.2576, 1.5711, 0.26172, 0.109, -0.059584) \,.
\end{equation*}

\subsection*{$\boldsymbol{m_{46}^- 4}$}

\begin{equation*}
{\cal R}_4 = -0.020509 \,, \quad  {\cal R}_6 = -0.053926 \,, \quad \eta_V= 1.3004 \,,
\end{equation*}
\begin{equation*}
\text{masses}^2 = (0.4783, 0.10213, 0.034824, 0.030265, -0.0066676) \,.
\end{equation*}

\subsection*{$\boldsymbol{m_{46}^- 5}$}

\begin{equation*}
{\cal R}_4 = -0.019001 \,, \quad  {\cal R}_6 = -0.072802 \,, \quad \eta_V=  1.2548 \,,
\end{equation*}
\begin{equation*}
\text{masses}^2 = (0.4181, 0.16632, 0.043898, 0.028584, -0.0059604) \,.
\end{equation*}

\end{appendix}

\newpage

\providecommand{\href}[2]{#2}\begingroup\raggedright
\endgroup

\end{document}